\documentclass[11pt,a4paper]{article}
\DeclareMathAlphabet{\scr}{U}{rsfs}{m}{n}

\usepackage{jheppub}

\usepackage{latexsym}
\usepackage{epsfig}
\usepackage[mathscr]{eucal}
\usepackage{amsfonts}
\usepackage{amscd}
\usepackage[capitalize]{cleveref}
\usepackage{array}   
\usepackage{amssymb}
\usepackage{colordvi}
\usepackage{enumerate}
\usepackage{graphicx}
\usepackage{booktabs}
\usepackage{theorem}
\usepackage{multirow}
\usepackage{hhline}
\usepackage{xspace}
\usepackage{bbm}
\usepackage{units}
\usepackage{comment}
 \usepackage{slashed}
\usepackage{color}

\let\newabs=\abs
\let\newnorm=\norm
\let\newbraket=\braket
\let\newpb=\pb
\usepackage{physics}
\usepackage{float}

\let\abs=\newabs
\let\norm=\newnorm
\let\braket=\newbraket
\let\pb=\newpb

\usepackage{mathtools}

\DeclarePairedDelimiter\abs{\lvert}{\rvert}%
\DeclarePairedDelimiter\norm{\lVert}{\rVert}%

\newcommand{\newc}{\newcommand}
\newc{\be}{\begin{equation}}
\newc{\ee}{\end{equation}}
\newc{\bea}{\begin{eqnarray}}
\newc{\eea}{\end{eqnarray}}
\newc{\ol}{\overline}
\newc{\wt}{\widetilde}
\newc{\bs}{\boldsymbol}
\newc{\m}{\mathcal}
\newc{\lan}{\langle}
\newc{\ra}{\rangle}
\newc{\pa}{\partial}

\newcommand{\non}{\nonumber} 
\newcommand{\crn}{\nonumber \\}
\newcommand{\beq}{\begin{eqnarray}} 
\newcommand{\eeq}{\end{eqnarray}} 
\newcommand{\bpmatrix}{\begin{pmatrix}}
\newcommand{\epmatrix}{\end{pmatrix}}
\newcommand{\ba}{\begin{array}}
\newcommand{\ea}{\end{array}}
\newcommand{\braket}[1]{\left(#1\right)}

\newcommand{\fr}{\frac}

\newcommand{\al}{\alpha}

\newcommand{\doublet}[2]{\begin{pmatrix} #1 \\ #2 \end{pmatrix}}
\renewcommand{\eqref}[1]{Eq.~(\ref{#1})}

\newcommand{\tab}[1]{Table~\ref{#1}}

\newcommand{\DRb}{\overline{\text{DR}}}

\newcommand{\OS}{\text{OS}}

\renewcommand{\Re}{\text{Re}\,}

\newcommand{\ew}{\text{EW}}

\newcommand{\ie}{{\it i.e.\;}}
\newcommand{\eg}{{\it e.g.\;}}
\newcommand{\cf}{{\textit{cf.}\;}}
\newcommand{\bc}{\begin{center}}
\newcommand{\ec}{\end{center}}

\newcommand{\gev}{~\text{GeV}}

\newcommand{\pb}{{~\text{pb}}}

\newcommand{\MSbar}{{\overline{\text{MS}}}}
\newcommand{\DRbar}{\overline{\text{DR}}}
\newcommand{\ti}{\tilde}
\newcommand{\msusy}{M_{\text{SUSY}} }

\newcommand{\s}{\newline \vspace*{-3.5mm}}

\usepackage{bm}
\allowdisplaybreaks

\newcommand{\NMSSMCALC}{{\tt NMSSMCALC}\xspace}
\newcommand{\NMSSMTools}{{\tt NMSSMTools}\xspace}
\newcommand{\SASP}{{\tt SARAH/SPheno}\xspace}
\newcommand{\FlexibleSUSY}{{\tt FlexibleSUSY}\xspace}

\newcommand{\sm}{{\text{SM}}}
\newcommand{\nmssm}{{\text{NMSSM}}}

\newcommand{\susy}{{\text{SUSY}}}
\newcommand{\orderqcd}{{{\cal O}(\alpha_t\alpha_s) }}
\newcommand{\orderew}{{{\cal O}((\alpha_t+\alpha_\lambda+\alpha_\kappa)^2) }}
\newcommand{\ordernew}{{{\cal O}(\alpha_{\text{new}}^2) }}

\newcommand{\Deltatwo}{\Delta^{ \text{\tiny(}\!\text{\tiny2}\!\text{\tiny)}}\!}
\newcommand{\Deltaone}{\Delta^{ \text{\tiny(}\!\text{\tiny1}\!\text{\tiny)}}\!}

\begin{document}
\title{
\vspace*{0cm}
\phantom{h} \hfill\mbox{\small } \\[-0.1cm]
\phantom{h} \hfill\mbox{\small KA-TP-17-2023}\\[-5mm]
\phantom{h} \hfill\mbox{\small DESY-23-112}
\\[2cm]
\textbf{The ${\cal O}(\alpha_t+\alpha_\lambda+\alpha_\kappa)^2$
  Correction to the $\rho$ Parameter and its Effect on the W Boson Mass Calculation in the Complex NMSSM\\[4mm]}}

\date{}

\author[a]{Thi Nhung Dao,}
\author[b]{Martin Gabelmann,}
\author[c]{Margarete M\"{u}hlleitner,}
 
\affiliation[a]{Faculty of Fundamental Sciences, PHENIKAA University, Hanoi 12116, Vietnam}
\affiliation[b]{Deutsches Elektronen-Synchrotron DESY, Notkestr.~85, 22607 Hamburg, Germany}
\affiliation[c]{Institute for Theoretical Physics, Karlsruhe Institute of Technology, Wolfgang-Gaede-Str. 1, 76131 Karlsruhe, Germany}
 
\emailAdd{nhung.daothi@phenikaa-uni.edu.vn}
\emailAdd{martin.gabelmann@desy.de}
\emailAdd{margarete.muehlleitner@kit.edu}
\abstract{
We present the prediction of the electroweak $\rho$ parameter and the $W$ boson
mass in the CP-violating Next-to-Minimal Supersymmetric extension of
the Standard Model
(NMSSM) at the two-loop order.
The $\rho$ parameter is calculated at the full one-loop and leading and
sub-leading two-loop order $\mathcal{O}(\alpha + \alpha_t\alpha_s +
\left(\alpha_t+\alpha_\lambda+\alpha_\kappa\right)^2)$.
The new $\Delta \rho$ prediction is incorporated into a prediction of $M_W$ via
a full supersymmetric (SUSY) one-loop calculation of $\Delta r$. Furthermore, we include all known state-of-the-art SM
higher-order corrections to $\Delta r$.
By comparing results for $\Delta \rho$ obtained using on-shell (OS) and $\DRbar$
renormalization conditions in the top/stop sector, we find that the scheme
uncertainty is reduced at one-loop order by 55\%, at two-loop
$\mathcal{O}(\alpha_s\alpha_t)$ by 22\%, and at two-loop 
$\mathcal{O}(\alpha_t+\alpha_\kappa+\alpha_\lambda)^2$ by 16\%, respectively.
The influence of the two-loop results on the $M_W$ mass prediction is found to
be sub-leading. The new calculation is made public in the computer program {\tt
NMSSMCALC}. We perform an extensive comparison in the $W$-mass, Higgs boson
mass and the muon anomalous magnetic moment prediction between our calculation
and three other publicly available tools and find very good agreement
provided that the  input parameters and
  renormalization scales are treated in the same way.
Finally, we study the impact of the CP-violating phases on the $W$-mass
prediction which is found to be smaller than the overall size of the SUSY
corrections.
}
\maketitle

\thispagestyle{empty}
\vfill
\newpage

\section{Introduction}
The Standard Model (SM) of particle physics has seen a tremendous
success story that certainly culminated in the discovery of the Higgs
boson in 2012 by the Large Hadron Collider (LHC) experiments ATLAS
\cite{Aad:2012tfa} and CMS \cite{Chatrchyan:2012ufa}. The theory has been tested extensively for its internal
consistency at the quantum level. In 2021, the combination of the
measurements of the $W$ boson mass has lead to a world average of
$M_W^{\text{exp}} =\unit[80.379\pm 0.012]{GeV}$
\cite{Zyla:2020zbs}. The SM predicts the $W$ boson mass to be $M_W^{\sm,\,\OS}=\unit[80.353\pm 0.004]{GeV}$ \cite{Bagnaschi:2022qhb} in the
on-shell (OS) renormalization scheme using the state-of-the-art
calculations \cite{Awramik:2003rn}, and to be
$M_W^{\sm,\MSbar}=\unit[80.351\pm 0.003]{GeV}$ in the $\MSbar$ scheme
\cite{Degrassi:2014sxa,Athron:2022isz} with the top mass being 172.76~GeV. This implies a deviation between the SM 
prediction of the $W$ boson mass and the experimental value of 
less than $2\sigma$ standard deviation, a tension which has been
existing between theory and experiment for a long time.
In 2022, the CDF collaboration has reported a new result of the 
 $W$ boson mass, which amounts to \cite{CDF:2022hxs}
\be 
M_W^{\text{CDF}} = \unit[80.4335\pm 0.0094]{GeV}.
\ee
The considerable shift of the central value and the small uncertainties of both
the individual CDF measurement and the SM prediction lead to
a discrepancy of more than $7\sigma$. 
Combining the new CDF result with the other
measurements from LEP, Tevatron and the LHC leads to a new world average
\cite{deBlas:2022hdk} of
\be 
M_W^{\text{exp}} = \unit[80.4133 \pm 0.0080]{GeV},
\ee
and a new deviation of order $6\sigma$. This result has caused a lot of
attention in the particle physics community. Many calculations have
been performed and analyzed in numerous models beyond the SM in order
to resolve this tension. Of particular interest are (minimal)
supersymmetric extensions of the SM which introduce 
fermionic/bosonic superpartners for each SM degree of freedom. In many
extensions, these superpartners carry an odd $R-$parity while the SM
particles carry an even $R-$parity. As a consequence of the imposed $R-$parity
conservation, only an even number of superpartners can contribute to
interaction vertices that also involve SM particles.
Therefore, amplitudes with only SM-like fields on external legs 
receive contributions from superpartners at most starting from the
one-loop order but not at tree level. 

The new particle content beyond the SM predicted by supersymmetry such
as \eg the superpartners of the SM fermions may give significant loop contributions to
the muon decay amplitude. Loop corrections to the muon decay are
usually parametrised with the quantity $\Delta r$ 
\cite{Sirlin:1980nh} defined through the matching of the Fermi theory and 
the high-energy theory. Subsequently, the experimentally well-known
muon life-time allows to exploit the relation between $\Delta r$, the $W/Z$ masses
$M_{W/Z}$, the Fermi constant $G_F$ and the fine-structure constant
$\alpha$ to make a precise prediction for $M_W$ in terms of the
other input parameters. Therefore, the combined $W$ boson mass
measurements can be used to constrain those
  parameters of  the
considered model that enter $\Delta r$ at a given loop level.
  
Recent studies \cite{Athron:2022isz,Bagnaschi:2022qhb,Yang:2022gvz} in
the Minimal Supersymmetric  
extension of the SM (MSSM) have shown that light and compressed
spectra of electroweakinos can yield $M_W^{\text{MSSM}}$ up to $\unit[80.376]{GeV}$ if taking into account 
the constraints from  current LHC supersymmetric (SUSY) searches and
limits on Dark Matter (DM) direct 
detection cross-sections and also satisfying the recent experimental result for the anomalous magnetic moment of the muon. In this work we consider the prediction of
$M_W$ within the Next-to-Minimal Supersymmetric Standard Model (NMSSM)
\cite{Fayet:1974pd,Barbieri:1982eh,Dine:1981rt,Nilles:1982dy,Frere:1983ag,Derendinger:1983bz,Ellis:1988er,Drees:1988fc,Ellwanger:1993xa,Ellwanger:1995ru,Ellwanger:1996gw,Elliott:1994ht,King:1995vk,Franke:1995tc,Maniatis:2009re,Ellwanger:2009dp}.
The model contains two Higgs doublet- and an additional complex
singlet-superfield. The $W$ boson mass in this model has been first studied in 
\cite{Domingo:2011uf} which includes the full one-loop corrections to
$\Delta r$.  The  authors of \cite{Stal:2015zca}
improved the calculation by taking into account not only the one-loop
but also the two-loop corrections of ${\cal O}(\alpha\alpha_s)$
\cite{Djouadi:1996pa,Djouadi:1998sq} and 
${\cal O}(\alpha_t^2,\alpha_t\alpha_b,\alpha_b^2)$
\cite{Heinemeyer:2002jq,Haestier:2005ja} to the  $\Delta \rho$ parameter into
account. 
It is important to stress that all contributions
to $\Delta \rho$ beyond one-loop order 
have been computed in the MSSM-limit so far. In this work, however, we
explicitly calculate $\Delta \rho$ including the full dependence on all
NMSSM-specific parameters at the two-loop level. 

Our group has contributed to the precise calculation of the Higgs boson masses in the 
real and complex NMSSM by computing the full one-loop
\cite{Ender:2011qh,Graf:2012hh}, and the 
two-loop ${\cal O}(\alpha_t \alpha_s)$ \cite{Muhlleitner:2014vsa}, ${\cal
O}(\alpha_t^2)$ \cite{Dao:2019qaz} as well as ${\cal
O}((\alpha_t+\alpha_\lambda+\alpha_\kappa)^2)$ \cite{Dao:2021khm} corrections. In
these two-loop computations the $W$ and $Z$ boson self-energies had to be
computed as a by-product at the corresponding orders.
Therefore, these results can be used to further improve the prediction
of the $W$ boson mass in the NMSSM by including not only the two-loop
QCD corrections but  also the two-loop electroweak ones of ${\cal
  O}((\alpha_t+\alpha_\lambda+\alpha_\kappa)^2)$, which contain
specific NMSSM corrections related to the NMSSM superpotential
parameters $\lambda$ and $\kappa$. 
The $\rho$ parameter and $W$ boson
mass calculations  are implemented in the Fortran code {\tt NMSSMCALC}
\cite{Baglio:2013iia,Ender:2011qh,Graf:2012hh,Muhlleitner:2014vsa,Dao:2019qaz,Dao:2021khm}
which is publicly available. The program computes the Higgs boson
masses and mixing angles up to two-loop ${\cal
  O}((\alpha_t+\alpha_\lambda+\alpha_\kappa)^2+\alpha_t \alpha_s)$,
together with the Higgs boson 
decay widths and branching ratios taking into account the most up-to-date
higher-order QCD corrections. The corrections to the trilinear Higgs
self-couplings are included at full one-loop level \cite{Nhung:2013lpa} and at
two-loop ${\cal O}(\alpha_t \alpha_s)$ \cite{Muhlleitner:2015dua} and
two-loop ${\cal O}(\alpha_t^2)$ \cite{Borschensky:2022pfc}. For the
CP-violating NMSSM the 
calculation of the electric dipole moments (EDMs) has been implemented \cite{King:2015oxa} to be
checked against the experimental constraints. Recently, the electron
and muon anomalous magnetic moments have been included in the code
\cite{Dao:2021vqp}. The code has been extended to include the
electroweak corrections to the NMSSM Higgs decays in {\tt NMSSMCALCEW}
\cite{Baglio:2019nlc,Dao:2019nxi,Dao:2020dfb}. 

This paper is organized as follows. In \cref{sec:tree-levelspectrum}, we briefly
introduce the used notation to describe the complex NMSSM as well as the
tree-level transformations to rotate from gauge to mass eigenstates.
The calculation of the one- and two-loop corrections to $\Delta\rho$
and the loop-corrected $W$ boson mass will be presented in 
 \cref{sec:rho} and \cref{sec:WMass}, respectively.
Furthermore, we discuss the renormalization schemes used to obtain UV-finite 
$\Delta\rho$ and $\Delta r$. \cref{sec:analysis} is dedicated to the
numerical analysis where we present the size of the loop corrections and
discuss their behaviour as a function of the NMSSM-specific parameters.
Finally, we give our conclusions in \cref{sec:conclusions}.

\section{The NMSSM at Tree Level}
\label{sec:tree-levelspectrum}
In this section, we give a short description of the complex NMSSM and introduce
the notation used in this paper. We follow the same notation which has been used in \cite{Graf:2012hh,Baglio:2013iia,Muhlleitner:2014vsa,Dao:2019qaz,Dao:2021khm}. 
The superpotential of the complex NMSSM is given by ($i,j=1,2$)
\begin{align}
    \label{eq:super}
    \mathcal{W}_{\nmssm} = 
	\epsilon_{ij} [y_e \hat{H}^i_d \hat{L}^j
\hat{E}^c + y_d \hat{H}_d^i \hat{Q}^j \hat{D}^c - y_u \hat{H}^i_u
\hat{Q}^j \hat{U}^c] -\epsilon_{ij} \lambda \hat S \hat H_d^i\hat H_u^j+\frac{1}{3}\kappa {\hat S}^3 \;,
\end{align}
with the quark and leptonic superfields $\hat{Q}$, $\hat{U}$, $\hat{D}$, $\hat{L}$, $\hat{E}$,
the Higgs doublet superfields $\hat{H}_d$, $\hat{H}_u$, the singlet superfield
$\hat{S}$ and the totally antisymmetric tensor $\epsilon_{12}= \epsilon^{12}=1$.
Charge conjugated fields are denoted by the superscript $c$.
Color and generation indices have been suppressed for the sake of clarity.
The Yukawa couplings $y_u,\,y_d$ and $y_e$ are assumed to be diagonal 3$\times$3 matrices in flavour space. 
The parameters $\lambda$ and $\kappa$ are in general complex.
The soft SUSY-breaking Lagrangian reads
\begin{align}\label{eq:breaking_term}\notag
\mathcal{L}_{\rm soft,NMSSM} = & -m_{H_d}^2 H_d^\dagger H_d - m_{H_u}^2
H_u^\dagger H_u -
m_{\tilde{Q}}^2 \tilde{Q}^\dagger \tilde{Q} - m_{\tilde{L}}^2 \tilde{L}^\dagger \tilde{L}
- m_{\tilde{u}_R}^2 \tilde{u}_R^*
\tilde{u}_R - m_{\tilde{d}_R}^2 \tilde{d}_R^* \tilde{d}_R
\nonumber      \\\nonumber
& - m_{\tilde{e}_R}^2 \tilde{e}_R^* \tilde{e}_R - (\epsilon_{ij} [y_e A_e H_d^i
\tilde{L}^j \tilde{e}_R^* + y_d
A_d H_d^i \tilde{Q}^j \tilde{d}_R^* - y_u A_u H_u^i \tilde{Q}^j
\tilde{u}_R^*] + \mathrm{h.c.})      \\
& -\frac{1}{2}(M_1 \tilde{B}\tilde{B} + M_2
\tilde{W}_j\tilde{W}_j + M_3 \tilde{G}\tilde{G} + \mathrm{h.c.}) \\ \nonumber
& - m_S^2 |S|^2 +
(\epsilon_{ij} \lambda
A_\lambda S H_d^i H_u^j - \frac{1}{3} \kappa
A_\kappa S^3 + \mathrm{h.c.}) \;,
\end{align}
where $H_{u,d}$, $S$, $\tilde{Q}$, $\tilde{L}$ and $\tilde{x}_R$ ($x=e,u,d$) are the scalar components of the respective superfields.
The fermionic superpartners $\tilde{B}$ (bino), $\tilde{W}_{1,2,3}$ (wino) and
$\tilde{G}$ (gluino) of the $U(1)_Y$, $SU(2)_L$ and $SU(3)_c$ gauge bosons obtain
the soft SUSY-breaking gaugino masses $M_1$, $M_2$ and $M_3$,
respectively. The mass parameters $M_{1,2,3}$ and the soft
SUSY-breaking trilinear couplings $A_x$ ($x=\lambda,\kappa,u,d,e$) are
in general complex, while the mass parameters of the scalar fields are real. 

After electroweak symmetry breaking, the Higgs boson fields are
expanded around their vacuum expectation values (VEVs) $v_u$, $v_d$,
and $v_s$, respectively,  
\begin{equation}
    H_d = \doublet{\frac{v_d + h_d +i a_d}{\sqrt 2}}{h_d^-}, \,\, 
    H_u = e^{i\varphi_u}\doublet{h_u^+}{\frac{v_u + h_u +i a_u}{\sqrt 2}},\,\,
    S= \frac{e^{i\varphi_s}}{\sqrt 2} (v_s + h_s + ia_s)\, ,
   \label{eq:vevs}
\end{equation}
with the CP-violating phases $\varphi_{u,s}$. The relation
to the SM VEV $v\approx 246.22$~GeV is given by
\begin{equation}
   v^2 = v_u^2 +v_d^2\,, 
\end{equation} 
and we define the mixing angle $\tan\beta$ as
\begin{equation}
   \tan\beta = \fr{v_u}{v_d} \;.
   \label{eq:tan_beta}
\end{equation}
Thus, the expressions for the tree-level weak
gauge boson masses are the
same as in the SM,
\be M_W^2= \fr14 g_2^2(v_u^2 +v_d^2), \quad  M_Z^2= \fr14 (g_1^2+g_2^2)(v_u^2 +v_d^2),\ee
where $g_1,g_2$ are the gauge couplings of the $U(1)_Y$ and $SU(2)_L$
gauge groups, respectively. These couplings can be written in terms of
the electric charge $e$ and the weak mixing angle $\theta_W$ as
\be  
g_1= \fr{e}{c_{\theta_W}},\quad g_2= \fr{e}{s_{\theta_W}}, 
\ee
 where the short hand notation $c_{x} \equiv \cos(x), s_{x} \equiv
 \sin(x), t_{x} \equiv \tan(x)$ is used in this paper.
The effective $\mu$ parameter is given by
\be 
\mu_{\text{eff}} = \fr{\lambda v_s e^{i\varphi_s}}{\sqrt 2}.
\ee 
Besides the particles of the SM,
gauge bosons, quarks, charged leptons, and three left-handed neutrino
fields, the NMSSM particle spectrum features an extended Higgs sector
and new SUSY particles, in particular: 
\begin{itemize}
\item 
The CP-even and CP-odd Higgs interaction states $(h_{d,u,s},
a_{u,d,s})$ mix to form five CP-indefinite Higgs mass eigenstates $h_{i}$ 
($i=1,...,5$), with their masses per convention ordered as   
$m_{h_1}\leq\dots\leq m_{h_5}$, and one neutral
Goldstone boson $G^0$. We use a two-fold rotation 
to rotate from the interaction to the mass eigenstates, 
\beq 
 (h_d, h_u,h_s, a, a_s,G^0)^T & = & \mathcal{R}^G(\beta)\, (h_d, h_u,
h_s, a_d, a_u, a_s)^T,\\
 (h_1,h_2,h_3,h_4,h_5,G^0)^T& = & \mathcal{R}^H\, (h_d, h_u,
h_s, a, a_s,G^0)^T,
\eeq
 where the first rotation matrix $\mathcal{R}^G$ with one rotation
 angle $\beta$ singles out the neutral Goldstone boson and the second
 rotation matrix $\mathcal{R}^H$ rotates the five interaction states
 $(h_d, h_u, h_s, a, a_s)$ to the five mass eigenstates
 $(h_1,h_2,h_3,h_4,h_5)$.
\item The charged Higgs interaction states $h_d^\pm,h_u^\pm$
  build up the charged Higgs
  bosons $H^\pm$ with  mass $M_{H^\pm}$
  and the charged Goldstone bosons $G^\pm$. 
\item The fermionic superpartners of the neutral Higgs bosons, the neutral
higgsinos $\tilde{H}_u$, $\tilde{H}_d$ and the singlino
$\tilde{S}$, mix with the neutral gauginos $\tilde{B}$ and
$\tilde{W}_3$, resulting in five neutralino mass eigenstates denoted as
$\tilde{\chi}^0_i$, $(i=1,...,5)$. 
The mass ordering of the  $\tilde{\chi}^0_i$ is chosen as 
$m_{\tilde{\chi}^0_1}\leq...\leq m_{\tilde{\chi}^0_5}$, and 
the rotation matrix $N$ transforms the fields $(\tilde{B},\ \tilde{W}_3,\ \tilde{H}_d,\ \tilde{H}_u,
\tilde{S})^T$  into the mass eigenstates.
\item The two chargino mass eigenstates,
\beq
\tilde{\chi}_i^+ = \left( 
    \begin{array}{c} 
        \ti{\chi}_{L_i}^+ \\[0.1cm]
        {\ti \chi_{R_i}^{-^*}}
    \end{array}\right) \;, \quad
                      i=1,2 \;,
\eeq
are obtained from the rotation of the interaction states, given by
  the charged Higgsinos $\tilde{H}^\pm_d$, $\tilde{H}^\pm_u$ and the
  charged gauginos $\tilde{W}^\pm=\left(\tilde{W}_1\mp i \tilde{W}_2\right)/\sqrt{2}$, to the mass eigenstates. This is
  achieved by using a bi-unitary transformation with the two $2\times 2$ unitary
  matrices $V^\chi$ and $U^\chi$,
 \begin{align}
 \ti{\chi}_L^+=V^{\chi}(\tilde{W}^+, \tilde{H}^+_u)^T,\ \ \
  \ti{\chi}_R^-=U^{\chi}(\tilde{W}^-, \tilde{H}^-_d)^T. 
 \end{align}
 \item The scalar partners of the left- and right-handed up-type
   quarks are denoted by $\ti u_{L/R}^i$, of the down-type quarks by $\ti
   d^i_{L/R}$, and of the charged leptons by $\ti l^i_{L/R}$
   ($i=1,2,3$). We do not include flavor mixing. Within each flavour
   the left- and right-handed scalar fermions with same electric
   charge mix. The sfermions are rotated to the mass eigenstates by a
   unitary matrix $U^{\ti f}$. 
   \item There are three scalar partners of the left-handed neutrinos,
       denoted as $\ti \nu_i$ ($i=1,2,3$) with the 
       masses
\be 
m^2_{\ti \nu_i} = \fr{1}2 M_Z^2 c_{2\beta} + m_{\ti{L}_i}^2 \;. 
\ee
\end{itemize}
For detailed discussions and derivations of the tree-level spectrum we refer
to earlier works
\cite{Graf:2012hh,Baglio:2013iia,Muhlleitner:2014vsa,Dao:2019qaz,Dao:2021khm}.
\section{The One- and Two-Loop Corrections to the $\rho$ Parameter }
\label{sec:rho}
In this section we describe the calculation of the full one-loop and the 
$\orderqcd$ and $\orderew$ two-loop corrections to the $\rho$ parameter.
The $\rho$ parameter is defined by the neutral- to charged-current coupling ratio at 
zero external momentum transfer \cite{Ross:1975fq,Veltman:1977kh,vanderBij:1986hy}
\be \rho = \fr{G_{NC}(0)}{G_{CC}(0)},\ee
where the neutral current coupling $G_{NC}(0)$ can be determined from, for
example, the coefficient of the $\nu_e \bar\nu_e\to \nu_e\bar\nu_e$ scattering
amplitude and the charged current coupling $G_{CC}(0)$ from the $e \bar\nu_e\to
e\bar\nu_e$ scattering amplitude. In the NMSSM, at tree level we have 
\be 
G_{NC}(0) = \fr{g_2^2}{2M_Z^2c_W^2}, \quad G_{CC}(0) =
\fr{g_2^2}{2M_W^2} ,
\ee
therefore $\rho^{(\text{tree-level})}=1$, as in the SM.  Higher-order
corrections to the $\rho$ parameter lead to a
deviation from unity, which can be written as
\cite{vanderBij:1986hy,Consoli:1989fg,Fleischer:1993ub}    
\bea 
    \rho &=& \fr{1}{1-\Delta \rho}= 1 + \Deltaone \rho + ((\Deltaone \rho)^2 +
  \Deltatwo \rho))+\cdots, \\
       \Delta \rho&=& \Deltaone \rho +\Deltatwo \rho+\cdots,
\eea
where the superscript $(n)$ indicates that the calculation is
performed at the $n$-loop order.  $\Delta^{(n)} \rho$ can be obtained
by computing $G_{CC}(0)$ and $G_{NC}(0) $ taking into account only
contributions related to the $W$ and $Z$ self-energies and 
expanding the ratio $G_{NC}(0)/G_{CC}(0)$ according to the considered loop
order. We get the following expressions for 
the one-loop and two-loop corrections 
\bea  
\Deltaone \rho &=& \fr{\Sigma_{ZZ}^{(1),T}}{M_Z^2}-\fr{\Sigma_{WW}^{(1),T}}{M_W^2},\\
\Deltatwo \rho &=& -\fr{\Sigma_{ZZ}^{(1),T}}{M_Z^2}  \braket{\fr{\Sigma_{ZZ}^{(1),T}}
      {M_Z^2}-\fr{\Sigma_{WW}^{(1),T}}{M_W^2}} +
    \braket{\fr{\Sigma_{ZZ}^{(2),T}}{M_Z^2}- 
      \fr{\Sigma_{WW}^{(2),T}}{M_W^2}},\label{eq:deltarhotwo}
\eea
where the transverse part of the gauge boson self-energy
$\Sigma_{VV}^{(n),T}$ ($V\equiv W,Z$)
is evaluated at zero external momentum. These expressions are the same
for the SM and for the 2-Higgs Doublet Model (2HDM) presented in
\cite{Hessenberger:2016atw} and hold in 
general for every theory with $\rho=1$ at tree level. 
In this paper, we compute the gauge boson self-energies at one- and
two-loop level taking into account the NMSSM particles and
couplings. These contributions also include the SM-like corrections as a subset.
When investigating the size of new physics effects, it is convenient to subtract the SM-like corrections.
To be consistent, the computation of the SM-like contributions
is performed in the same way as the NMSSM-specific contributions.
This strategy can be followed when considering $\Delta \rho$.
In the investigation of the $W$ boson mass it is, however, crucial to include existing
SM-like higher-order corrections to $\Delta r$ beyond the two-loop
level (\cf \cref{sec:WMass} for a detailed discussion).
\\
\underline{\bf{The one-loop result}}\\
The transverse parts of the gauge boson self-energies are computed with the full
content of the NMSSM at zero external momentum without any further approximations.
In order not to break SUSY at loop-level, we perform the calculation using
dimensional reduction ($\DRb$) rather than minimal subtraction ($\MSbar$).
While the results for the individual gauge boson
self-energies differ when using $\MSbar$ or $\DRbar$, we find that the difference
of the $W$ and $Z$ self-energies, entering $\Delta \rho$, is the same
in both regularization schemes. We confirmed this explicitly at the
one- and two-loop level. This is considered as a further consistency
check of the results. 

We can decompose the $ \Deltaone \rho$ into four contributions given
by the SM fermion, the Higgs, the squark and slepton, and the chargino
and neutralino ones, 
\be 
\Deltaone \rho = \Delta^{f} \rho+\Delta^H \rho
+\Delta^{\ti f}  \rho 
+\Delta^{\ti\chi} \rho \,.
\ee   
They are separately UV-finite. The SM fermion contributions are the
same as in the SM. The contributions from the first  
two generations can be neglected, and the contribution from the
third generation reads
\be 
\Delta^{f} \rho = \fr{m_\tau^2 + 3 m_b^2+3 m_t^2 }{16\pi^2 v^2} +
\fr{3 m_b^2m_t^2 \ln\fr{m_t^2}{m_b^2}}{8 \pi^2v^2 (m_b^2-m_t^2)} \,.
\ee
The contribution from the Higgs sector, \ie from $G^\pm, G^0, h^i$ and $H^\pm$, reads
\begin{align}
\Delta^H \rho = &\fr{1}{8v^2\pi^2}\bigg[ s_{2\theta_W}^2  A_0(M_{H^\pm}^2 )
- 2 c_{2\theta_W}^2 B_{00}(M_{H^\pm}^2 ,M_{H^\pm}^2 )- \sum_{i,j=1}^5 4g_{h_ih_jZ}^2B_{00}(
 m_{h_i}^2  ,m_{h_j}^2  )  \crn
 & +\sum_{i=1}^5\left( 8 \abs{g_{h_iH^-W^+}}^2 B_{00}(m_{h_i}^2,M_{H^\pm}^2 )
  + 8 \abs{g_{h_iG^-W^+}}^2 B_{00}(m_{h_i}^2,0 )\right)\bigg]
 \,,
 \end{align}
where the one-loop functions are
given at the end of this section and
\begin{align}
g_{h_iH^-W^+}& = \fr12 (-{\cal R}^H_{i4} - i(s_\beta {\cal R}^H_{i1} -c_\beta {\cal R}^H_{i2 }))\,, \quad  
 g_{h_iG^-W^+} = \fr12   i(c_\beta {\cal R}^H_{i1} +s_\beta {\cal R}^H_{i2 }) \,,\quad\crn
g_{h_ih_jZ}&=\fr12 \left( {\cal R}^H_{i4}(s_\beta {\cal R}^H_{j1} -c_\beta {\cal R}^H_{j2}
             )- {\cal R}^H_{j4}(s_\beta {\cal R}^H_{i1} -c_\beta {\cal R}^H_{i2} )  \right) \;,
\end{align}
with $c_{2W} \equiv \cos 2 \theta_W, s_{2W} \equiv \sin 2
\theta_W$. 

The contribution from the third generation squarks
  and sleptons is given by 
\begin{align}
\Delta^{\ti f} \rho =&\fr{1}{8v^2\pi^2}\bigg[-2 
\sum_{i=1,2}\sum_{\ti f=\ti \tau,\ti t,\ti b}n_{\ti f}Q_{\ti f_i}^2s_{\theta_W}^2\left[n_{\ti f} (n_{\ti f} \abs{ Q_{\ti f_i}}-  2 t_{\ti f_i} ) \abs{U_{i1}^{\ti f}}^2  - 2 s_{\theta_W}^2
 \right] A_0(m_{\ti f_i}^2)
                       \crn
&-2 B_{00}(m_{\ti \nu_3}^2,m_{\ti \nu_3}^2) - 2\sum_{i=1,2}\left[ \abs{ U_{i1}^{\ti \tau} }^2  - 2 s_{\theta_W}^2\right] ^2  B_{00}(m_{\ti \tau_i}^2,m_{\ti \tau_j}^2)\crn
& - \fr 23 \sum_{i=1,2}\sum_{\ti f=\ti t,\ti b} n_{\ti f}^2\left[ \abs{
  U_{i1}^{\ti f} }^2  - 2\abs{Q_{\ti f_i}} s_{\theta_W}^2\right] ^2  B_{00}(m_{\ti f_i}^2,m_{\ti f_j}^2)\crn 
& + 4\sum_{i=1,2} \abs{ U_{i1}^{\ti \tau} }^2B_{00}(m_{\ti \tau_i}^2,m_{\ti \nu_3}^2) + 4n_{\ti f}
\sum_{i,j=1,2} \abs{ U_{i1}^{\ti t} }^2 \abs{ U_{i1}^{\ti b} }^2 B_{00}(m_{\ti b_i}^2,m_{\ti t_j}^2)
\bigg]\,,
\end{align}
where  $n_{\ti f}=3$ for squarks and $n_{\ti f}=1$ for sleptons, 
 $t_{\ti \tau_i/\ti b_i}=-1/2, t_{\ti t_i}=1/2$, $Q_{\ti f_i}$
 denotes the electric charge of the sfermions. 
The contribution from the charginos and neutralinos can be cast into
the form 
\begin{align}
\Delta^{\ti \chi} \rho =
  &\fr{1}{2v^2\pi^2}\bigg\{\sum_{i,j=1}^5 \bigg[  \Re{(g_{\ti\chi^0_i\ti\chi^0_j Z}^{L} g_{\ti\chi^0_j\ti\chi^0_i Z}^{R})} 
    m_{\ti
    \chi_i^0}m_{\ti
    \chi_j^0}   B_0 (m^2_{\ti\chi_i^0},m^2_{\ti\chi_j^0}) \crn
  &  
  -\fr12  (\abs{g_{\ti\chi^0_i\ti\chi^0_j Z}^{L}}^2+\abs{g_{\ti\chi^0_i\ti\chi^0_j Z}^{R}}^2)
 {\cal F} (m^2_{\ti\chi_i^0}, m^2_{\ti\chi_j^0}) \bigg] \crn
  & + \sum_{i=1}^2 \sum_{j=1}^5 \,\bigg[
- 2 \Re{ (g^L_{\ti\chi_i^\pm\ti\chi_j^0
    W}(g^R_{\ti\chi_i^\pm\ti\chi_j^0 W})^*  )} m_{\ti \chi_i^\pm}
    m_{\ti \chi_j^0} 
 B_0 (m^2_{\ti\chi_i^\pm},m^2_{\ti\chi_j^0})  \crn
 &+ (\abs{g^L_{\ti\chi_i^\pm\ti\chi_j^0 W}}^2 +
  \abs{g^R_{\ti\chi_i^\pm \ti\chi_j^0 W}}^2 )    {\cal F}
 (m^2_{\ti\chi_i^\pm},m^2_{\ti\chi_j^0}) \bigg] \crn
 &+\sum_{i,j=1}^2 \,\bigg[ 2 \Re{( g^L_{\ti\chi_i^\pm\ti\chi_j^\pm
    Z}(g^R_{\ti\chi_i^\pm\ti\chi_j^\pm Z})^* ) } m_{\ti \chi_i^\pm}
    m_{\ti \chi_j^\pm} B_0 (m^2_{\ti\chi_i^\pm},m^2_{\ti\chi_j^\pm})   
  \crn &  +   (
 \abs{ g^L_{\ti\chi_i^\pm \ti\chi_j^\pm Z}}^2+\abs{g^R_{\ti\chi_i^\pm\ti\chi_j^\pm Z}  }^2) 
  {\cal F} (m^2_{\ti\chi_i^\pm}, m^2_{\ti\chi_j^\pm})  \bigg] \bigg\} \,,
\end{align}
where 
\begin{align}
g_{\ti\chi^0_i\ti\chi^0_j Z}^{L}&=-\fr12 (N_{i3}N_{j3}^*-N_{i4}N_{j4}^*), \quad &g_{\ti\chi^0_i\ti\chi^0_j Z}^{R}&=-g_{\ti\chi^0_j\ti\chi^0_i Z}^{L},\crn
g^L_{\ti\chi_i^\pm\ti\chi_j^\pm
    Z}&= c_{\theta_W}^2 V^{\chi}_{i1}V^{\chi*}_{j1}+\fr12 c_{2\theta_W} V^{\chi}_{i2}V^{\chi*}_{j2}, \quad 
   & g^R_{\ti\chi_i^\pm\ti\chi_j^\pm
    Z}&= c_{\theta_W}^2 U^{\chi}_{j1}U^{\chi*}_{i1}+\fr12 c_{2\theta_W} U^{\chi}_{j2}U^{\chi*}_{i2},\crn
 g^L_{\ti\chi_i^\pm\ti\chi_j^0
    W} &= V^{\chi}_{i1} N_{j2}^* + \fr{1}{\sqrt{2}}  V^{\chi}_{i2} N_{j3}^*, \quad &
     g^R_{\ti\chi_i^\pm\ti\chi_j^0
    W} &= U^{\chi}_{i1} N_{j2} - \fr{1}{\sqrt{2}}  U^{\chi}_{i2} N_{j4}, \crn
{\cal F} (x, y) &= A_0(y) - 2 B_{00}(x,y) + y B_0(x,y).
\end{align}
In the above expressions, the one-loop integrals are defined as
\begin{align}
A_0(x) &= x (1 - \ln(\bar x)) \crn
B_0(x,y) &= \begin{cases}(1 +\fr{- x\ln(\bar x) + y\ln(\bar y)}{x-y}  ), \quad &x\ne y\\
                        -\ln(\bar x), \quad &x= y \end{cases} \crn
B_{00}(x,y) &= \begin{cases}\fr38 (x+y)+ \fr{- x^2\ln(\bar x) + y^2\ln(\bar y)}{4(x-y)} , \quad& x\ne y\\
                     \fr x2 (1 - \ln(\bar x)) , \quad & x= y \end{cases} ,
\end{align}
where the bar symbol indicates the dimensionless quantities, \ie
$\bar{x}=x/\mu_R^2$ etc.~with $\mu_R$ being the renormalization scale.
\\

\noindent \underline{\bf{The two-loop $\orderqcd$ corrections}}\\
For the $\orderqcd$ corrections, the first term in
\eqref{eq:deltarhotwo} will vanish and hence
\bea 
\Delta^{\alpha_t\alpha_s} \rho &=& \fr{\Sigma_{ZZ}^{\alpha_t\alpha_s,T}(0)}{M_Z^2}-
      \fr{\Sigma_{WW}^{\alpha_t\alpha_s,T}(0)}{M_W^2}. 
\eea
 We take the results of $\Sigma_{ZZ}^{\alpha_t\alpha_s,T}(0)$ and
  $\Sigma_{WW}^{\alpha_t\alpha_s,T}(0)$ implemented in {\tt NMSSMCALC}, which have been 
  computed by our group in Ref.~\cite{Muhlleitner:2015dua} in the complex NMSSM.
 For the detailed calculation we refer the reader to \cite{Muhlleitner:2015dua}, 
 here we summarize only the main features. We use the gaugeless limit,
 \ie we set the electric charge and the $W$ and $Z$ boson masses to zero, 
\begin{equation}
    \label{eq:gaugeless}
    e,M_W,M_Z\to 0,\,\, \frac{M_W}{M_Z} = \text{const.}
\end{equation}
The ratio $\fr{\Sigma_{VV}^{\alpha_t\alpha_s,T}(0)}{M_V^2}$ ($V=W,Z$),
however, is non-zero and proportional to $\alpha_t\alpha_s$. 
In our calculation, we set the bottom quark mass to zero. The two-loop $\orderqcd$ 
corrections can be decomposed into the contributions from the genuine two-loop
diagrams (containing either a gluon/gluino loop or a loop with a
stop/sbottom quartic coupling)
and the contributions from the counterterm inserted one-loop diagrams (containing 
either coupling-type counterterms or propagator-type counterterms). 
The set of independent parameters entering the top/stop and bottom/sbottom sector, 
that need to be renormalized at ${\cal O}(\alpha_s)$ are
\be 
m_t, \; m_{\tilde{Q}_3}, \; m_{\tilde t_R} \quad \mbox{and} \quad A_t \;.
\label{eq:stopparset}
\ee
In \cite{Muhlleitner:2015dua} we have discussed two renormalization
schemes for these parameters, on-shell (OS) and $\DRbar$ renormalization. We keep
these two options of renormalization schemes here, too. 
The expressions for the required counterterms were presented in
\cite{Muhlleitner:2015dua}. 
We have explicitly confirmed that the obtained result for
$\Delta^{\alpha_t\alpha_s}_{\nmssm} \rho$ is UV finite. 
For the SM-like contributions we reproduce the known result in the literature
\cite{Djouadi:1987gn,Djouadi:1987di},  
\be 
\Delta^{\alpha_t\alpha_s}_{\sm} \rho =- (1+ \pi^2/3)\fr{\al_s m_t^2}{
(8 \pi^3 v^2)} ,
\ee 
where the top mass is renormalized using the OS renormalization scheme.
Note that all our calculations have been 
performed in dimensional reduction for both the SM and the NMSSM, while the results in
\cite{Djouadi:1987gn,Djouadi:1987di} were obtained using dimensional
regularization. 
It should be stressed that $\Delta^{\alpha_t\alpha_s} \rho$ in the NMSSM
is identical to the corresponding quantity in the MSSM
which has been calculated in \cite{Djouadi:1996pa,Djouadi:1998sq}. \s
\\
\underline{\bf{The two-loop $\orderew$ corrections}}\\
For the sake of a convenient notation we denote 
\be 
\alpha_{\ew}^2 \equiv (\alpha_t+\alpha_\lambda+\alpha_\kappa)^2 .
\ee 
Since the electroweak sector  contributes to the gauge boson
self-energies at one-loop level, its contribution to $\Delta\rho$ at
two-loop level contains also the one-loop squared terms, hence
\bea
    \Delta^{\alpha_{\ew}^2}  \rho &=& -\fr{\Sigma_{ZZ}^{(1),T}}{M_Z^2}
    \braket{\fr{\Sigma_{ZZ}^{(1),T}} 
    {M_Z^2}-\fr{\Sigma_{WW}^{(1),T}}{M_W^2}} +
  \braket{\fr{\Sigma_{ZZ}^{(2),T}}{M_Z^2}- 
    \fr{\Sigma_{WW}^{(2),T}}{M_W^2}}.\label{eq:deltarhoew}
\eea
The calculation of the transverse part of the gauge boson self-energies at
the one- and two-loop order in the complex NMSSM has 
been  presented in \cite{Dao:2021khm} and implemented in  {\tt NMSSMCALC}. 
The results have been obtained in the gaugeless limit, \cf \cref{eq:gaugeless}. In this
limit the Higgs-Goldstone couplings are non-zero at
$\order{\alpha_\lambda+\alpha_\kappa}$ while the Goldstone tree-level masses
vanish which leads to intermediate infra-red (IR) divergences in some of the
two-loop diagrams that cancel in the sum of all self-energy
diagrams. 
The top/stop, bottom/sbottom, Higgsino/singlino and Higgs sectors contribute already at one-loop level. 
The two-loop self-energies $\Sigma_{VV}^{(2),T}$ consist of
contributions from the genuine two-loop diagrams 
and the counterterm inserted one-loop diagrams. Therefore
the parameters of these sectors need to be renormalized at one-loop
level to compute the two-loop self-energies $\Sigma_{VV}^{(2),T}$. For
the parameters of the Higgsino/singlino and Higgs sectors we  
apply a mixed $\DRb$-OS renormalisation
scheme while for the top/stop sector we apply an OS scheme or a $\DRbar$ scheme.
All counterterms for the complex phases
$\varphi_\alpha$ ($\alpha=s,u,\kappa,\lambda$) can be set to
zero. The remaining input parameters  together with the applied
renormalization conditions are given by 
\begin{eqnarray}
\underbrace{M_{H^\pm}^2,v,s_{\theta_W},}_{\mbox{OS
 scheme}} \underbrace{m_t,  m_{\tilde{Q}_3},  m_{\tilde t_R},
 A_t }_{\mbox{OS}/\overline{\mbox{DR}} \mbox{
 scheme}},
\underbrace{\tan\beta,|\lambda|,v_s,|\kappa|,\mbox{Re} A_\kappa}_{\overline{\mbox{DR}} \mbox{ scheme}}\,,
\label{eq:mixedcond1}
\end{eqnarray}
in case $M_{H^\pm}^2$ is used as independent input, or
\begin{eqnarray}
\underbrace{v,s_{\theta_W},}_{\mbox{OS
 scheme}} \underbrace{m_t,  m_{\tilde{Q}_3},  m_{\tilde t_R},
 A_t }_{\mbox{OS}/\overline{\mbox{DR}} \mbox{
 scheme}},
\underbrace{\tan\beta,|\lambda|,v_s,|\kappa|,\mbox{Re}
  A_\lambda,\mbox{Re} A_\kappa}_{\overline{\mbox{DR}} \mbox{ scheme}}\,,
\label{eq:mixedcond2}
\end{eqnarray}
if $\mbox{Re}A_\lambda$ is chosen as independent input rather than $M_{H^\pm}$.
The tadpole counterterms are defined such that the minima of the Higgs
potential do not change at higher order. We explicitly confirmed that $\Delta^{\alpha_{\ew}^2}  \rho$ is UV 
finite and is free of IR divergences (\cf
Ref.~\cite{Dao:2021khm} for a detailed discussion on the cancellation of all IR
divergences).

The corresponding result $\Delta^{\alpha_{\ew}^2}_{\text{SM}} \rho$
within the SM, which is subtracted from the NMSSM result, is computed
in the same way as described above, \ie using OS conditions for $v,\,\sin\theta_W$
and OS/$\DRbar$ conditions for $m_t$.
In the SM only contributions from top and Higgs diagrams enter at
$\order{\alpha_{\ew}^2}$, hence
  $\Delta^{\alpha_{\ew}^2}_{\text{SM}}  \rho = \Delta^{\alpha_{t}^2}_{\text{SM}} 
\rho$.
It should be noted that $\Delta^{\alpha_{\ew}^2}_{\text{SM}}  \rho$ does not depend on the renormalization of the Higgs mass
and the Higgs tadpole. Note that our expression of $\Delta^{\alpha_{t}^2}_{\text{SM}} 
\rho$ in \eqref{eq:deltarhoew} looks different from the one computed by Fleischer, Tarasov and Jegerlehner
in \cite{Fleischer:1993ub} for the following reasons.
In their computation, $\Delta\rho$ at two-loop order is defined as
\be 
\Delta^{\alpha_{t}^2}_{\text{SM, FTJ}}
 \rho=\braket{\fr{\Sigma_{ZZ}^{\alpha_t^2,T}(0)}{M_Z^2}-
      \fr{\Sigma_{WW}^{\alpha_t^2,T}(0)}{M_W^2}},\label{eq:FTJ}
\ee
where $\Sigma_{VV}^{\alpha_t^2,T}(0)$ gets contributions only from the genuine
two-loop diagrams and the counterterm inserted diagrams, \ie the one-loop-squared
pieces from Eq.~(\ref{eq:deltarhotwo}) are not contributing. This is because
they used the Fermi constant $G_\mu$ as an input parameter. Therefore,
in \cite{Fleischer:1993ub} only the top  
mass needs to be renormalized in order to obtain a UV finite result. 
However, in the following, we use the VEV $v$ (defined through $M_W$, $M_Z$
and $\alpha(0)$) as an input. Considering the one-loop expression
for $\Delta\rho$ in the SM, which involves both $m_t$ and $v$, 
the two-loop result needs to include the renormalization of $m_t$ and 
$v$. Since $v$ is renormalized in the OS scheme, its counterterm contributes with
a non-zero finite part of $\delta v$
which is not present in the result of \eqref{eq:FTJ}. The
UV-finite part of $\delta v$ also gives rise to non-vanishing contributions to
the single pole of the counterterm inserted one-loop diagrams. This additional
contribution is precisely canceled by the one-loop-squared term in
\eqref{eq:deltarhoew}. Our final results for $\Delta^{\alpha_{t}^2}_{\text{SM}} 
\rho$  are the same as the results of \cite{Fleischer:1993ub} provided that
 $v=1/ \sqrt{\sqrt{2}G_\mu}$.
The Higgs mass in the SM result $\Delta^{\alpha_{\ew}^2}_{\text{SM}}
\rho$ is set to be equal  
to the possibly non-zero tree-level SM-like Higgs boson mass in
$\Delta^{\alpha_{\ew}^2}_{\text{NMSSM}}  \rho$ 
obtained in the gaugeless limit. This is a crucial difference to the MSSM, where
the SM-like Higgs boson mass in the gaugeless limit always vanishes. 
\section{Calculation of the $W$ Boson Mass in the OS Scheme}
\label{sec:WMass}
The $W$ boson mass can be computed from the following relation \cite{Sirlin:1980nh},
\be 
\fr{G_\mu}{\sqrt{2}} = \fr{\pi \alpha }{2M_W^2 s_W^2} (1+\Delta
r), \label{eq:MWcalc} 
\ee 
where $G_\mu$ is the Fermi constant,
$\alpha\equiv \alpha (0)= e^2/4\pi$ is the fine-structure constant in the Thomson limit,
and $\Delta r$ includes all loop contributions
to the amplitude of the $\mu\to  e\bar\nu_e \nu_\mu$ decay after
subtracting the Fermi-model-type QED correction.
By using the OS weak mixing angle $s_W^2= 1- M_W^2/M_Z^2$,
\cref{eq:MWcalc} can be solved for $M_W$, 
\be 
M_W^2 = \fr{M_Z^2}{2}\left\{ 1+ \sqrt{1-  \fr{4\pi\alpha}{\sqrt{2}G_\mu M_Z^2}
        \left(1+\Delta^{(n)}_{\text{\tt NMSSMCALC}} r\right)
}  \right\} , \label{eq:MWcalcit}
\ee
where the NMSSM $\Delta r$ is taken from our implementation
  in {\tt NMSSMCALC}.
The quantity $\Delta^{(n)}_{\text{\tt NMSSMCALC}} r$, the subscript $(n)$ denotes the $n$-loop order, depends also on
$M_W$.
Therefore, \cref{eq:MWcalcit} has to be evaluated iteratively. 
In the following subsections we describe in detail the NMSSM-specific one-
and two-loop contributions as well as the higher-order SM corrections
which are included in $\Delta r$. 
\subsection{One-loop Corrections }
The one-loop correction to $\Delta r$ can be written as
\begin{align}
\Delta^{(1)} r&= \fr{\Sigma_T^{WW}(0)-\delta M_W^2}{M_W^2} +2\delta Z_e-
2 \fr{\delta s_W}{s_W} +\fr12 \left(\delta Z^\mu + \delta Z^e+
\delta Z^{\nu_\mu}+\delta Z^{\nu_e} \right)\crn
                   & + \Delta r_{\triangle}  + \Delta r_{\Box} \,, \label{eq:deltar}
\end{align} 
where the origin of the individual terms in this
expression is explained in the
following. Using the OS renormalization scheme for the  
electric charge, its counterterm is given by
\be 
\delta Z_e = \fr12 \Pi^{AA}(0) - \text{sgn} 
\fr{s_W}{c_W} \fr{\Sigma_T^{AZ}(0)}{M_Z^2} \;, \label{eq:deltaZe}  
\ee
where 
\be
\Pi^{AA}(0)\equiv\fr{ \Sigma_T^{AA}(k^2)}{ k^2}\bigg|_{k^2=0}
\ee 
and
$\text{sgn}=1$ for the SM and $\text{sgn}=-1$ for the NMSSM accounts for
different sign conventions in the covariant derivatives of the two models. 
The photon self-energy in \eqref{eq:deltaZe} evaluated at vanishing external
momentum contains large logarithmic contributions, $\order{\ln(\mu_R^2/m_q^2)}$,
which depend on ratios of the light quark masses and the renormalization scale
$\mu_R$, leading to numerical instabilities.
To avoid this dependence, one writes for the light SM fermions ($f=u,d,s,c,b,e,\mu,\tau$),
\bea
\Pi^{AA}(0)&=& \Pi^{AA}_f(0) - \Re\Pi^{AA}_{f}(M_Z^2)+\Re\Pi^{AA}_{f}(M_Z^2) 
 +\Pi^{AA}_{\text{rem}}(0)\crn
  &=& \Delta\alpha +
  \Re\Pi^{AA}_{f}(M_Z^2)+\Pi^{AA}_{\text{rem}}(0),
\eea 
where $\Re\Pi^{AA}_{f}(M_Z^2)$ is computed perturbatively,
$\Pi^{AA}_{\text{rem}}(0)$ contains contributions from remaining
charged particles of the model and
$\Delta\alpha=\Delta\alpha_{\text{lepton}} + 
\Delta\alpha_{\text{had}}^{(5)}$. The contribution $\Delta\alpha_{\text{had}}^{(5)}$
is determined from the dispersion relation using experimental data. We take 
$\Delta\alpha_{\text{had}}^{(5)}= 0.02768$ from Ref.~\cite{ParticleDataGroup:2022pth}. 
The quantity $\Delta\alpha_{\text{lepton}} = 0.03150$ includes contributions up to
three loops \cite{Steinhauser:1998rq}.  The counterterm of $s_W$ is derived from the OS 
relation $s_W^2= 1- M_W^2/M_Z^2$ and reads
\be
\delta s_W = \fr{c_W^2}{2 s_W}\left[ \fr{\delta M_Z^2}{M_Z^2} - \fr{\delta M_W^2}{M_W^2} \right],
\ee
where the $Z$ and $W$ boson mass counterterms $\delta M_Z^2$ and $\delta M_W^2$ are defined
in the OS scheme,
\be
\delta M_Z^2 =\Sigma^T_{ZZ}(M_Z^2) \;, \quad  \delta M_W^2
=\Sigma^T_{WW}(M_W^2) \;,
\ee
such that $M_W$ and $M_Z$ correspond to the real part of the complex pole of
the $W$ and $Z$ propagator with a constant decay width $\Gamma_W$ and
$\Gamma_Z$,
respectively. However, experimentally the resonances are parametrized using a
Breit-Wigner line shape that features an energy-dependent rather than fixed
decay width. To account for this, we convert the vector boson
masses that correspond to the standard OS definition to the running-width masses
 $M^{\text{run}}_{W/Z}= M_{W/Z} + \Gamma_{W/Z}^2/(2M_{W/Z}^{\text{run}})$ as
described in \cite{Bardin:1988xt}.
For $\Gamma_Z$ we use the experimentally measured value while for the $W$ boson width
we use the approximate formula $\Gamma_W=3G_\mu
M_W^3/(2\sqrt{2}\pi)(1+2\alpha_s/(3\pi))$ \cite{Freitas:2002ja}
neglecting possible beyond-the-SM (BSM)
contributions. While the conversion from $M_Z^{\text{run}}$ to $M_Z$ needs to be
done only once before the iteration in \cref{eq:MWcalcit}, the conversion from
$M_W$ to $M_W^{\text{run}}$ needs to be done at the end of the iteration. For
more details on the running/fixed-width mass conversion we refer
to \cite{Freitas:2002ja,Stal:2015zca}.

The wave function (WF) counterterms of the external
leptons, $\delta Z^l$, and the triangle and box contributions, $\Delta
r_{\triangle,\Box}$, are evaluated for vanishing lepton masses.
The WF counterterms of the external leptons are defined in the OS scheme,
\be 
\delta Z^l =-\Sigma_{ll}^L(0), \quad l=\mu,e,\nu_\mu,\nu_e \;,
\ee
where $\Sigma_{ll}^L(0)$ is the form-factor of the left-handed vector component of the lepton self-energy.
The Fermi-model type QED correction appears in $\delta Z^l$, the
triangle and the box diagrams. In order to remove this contribution, one should replace
the photon propagator in the loop diagrams of the WF counterterms by
\be
\fr{1}{k^2} \to \fr{1}{k^2 -M_W^2},
\ee 
and remove the box diagrams with virtual photons. 
The  vertex diagrams with virtual photons are calculated as usual. 
This procedure has been first introduced in Refs.~\cite{Sirlin:1980nh,Sirlin:1977sv}. 
Using this trick we recover the well-known relation between triangle, box and vertex
corrections for the SM, derived \eg in 
\cite{Denner:1991kt},
\bea 
\delta_{vb} &\equiv&\fr 12 (\delta Z^\mu+ \delta Z^e +
\delta Z^{\nu_\mu}+\delta Z^{\nu_e}) + \Delta_{\triangle} r + \Delta_{\Box} r - 
\fr{2 }{c_W s_W } \fr{\Sigma^T_{AZ}(0)}{M_Z^2} \,,\\
\delta_{vb}^{\text{SM}}  &=& \frac{\alpha}{4\pi s_W^2} \left( 6 + \fr{7-4s_W^2}{2s_W^2}\log c_W^2\right).
\eea
Collecting all previous derivations leads to the known SM result
\begin{align}
\Delta^{(\alpha)}_{\sm} r=&\Delta\alpha -
 \fr{c_W^2}{s^2_W}\left[ \fr{\delta M_Z^2}{M_Z^2} - \fr{\delta M_W^2}{M_W^2} \right] \\
 & 
 + \Re\Pi^{AA}_{f}(M_Z^2)+\Pi^{AA}_{\text{rem}}(0) + \fr{2c_W}{s_W} \fr{\Sigma^T_{AZ}(0)}{M_Z^2} +\fr{\Sigma^T_{WW}(0)-\delta M_W^2}{M_W^2}+ \delta_{vb}^{\sm} ,\nonumber
\end{align} 
which can be written as
\begin{align}
\Delta^{(\alpha)}_{\sm} r= \Delta\alpha -
 \fr{c_W^2}{s^2_W} \Delta^{(1)} \rho + \Delta r_{\text{rem}},
 \end{align}
 where $\Delta \alpha$ and $\Delta^{(1)} \rho$ contain the numerically dominant part of
the corrections.

Within the complex NMSSM, we have computed the one-loop contributions to
$\Delta^{(1)}_{\nmssm} r$ using the expression in \eqref{eq:deltar} and using the
resummation of the light fermions in the photon self-energy described above.
We have checked that the resulting expression is UV-finite and renormalization scale independent.
Contributions from squarks, sleptons, charginos and
neutralinos form separate UV-finite subsets and can be studied independently. 
On the other hand, contributions from gauge and Higgs bosons must be combined in
order to obtain UV-finite results.

\subsection{Combination with Known Higher-Order Corrections} 
\label{sec:smho}
In the SM, corrections $\Delta_{\sm}^{\text{lit.}} r$ up to four-loop order have been
computed and are available in the literature. Following the procedure developed for
the MSSM \cite{Heinemeyer:2013dia} and implemented in {\tt FeynHiggs} \cite{Bahl:2018qog,Bahl:2017aev,Bahl:2016brp,Hahn:2013ria,Frank:2006yh,Degrassi:2002fi,Heinemeyer:1998np,Heinemeyer:1998yj}, we have
combined all available higher-order SM corrections with the presented
NMSSM contributions. 
For fixed $n$ ($n=1,2$) we first subtract $\Delta^{\,(n)}_{\sm} r$ from $\Delta^{(n)}_{\nmssm}
r$ and re-add $\Delta_{\sm}^{\text{lit.}} r$ including all known-higher corrections from the
literature, 
\be
\Delta^{(n)}_{\susy}
r=\Delta^{(n)}_{\nmssm} r - \Delta^{(n)}_{\sm} r \,, 
\label{eq:deltar1}
\ee
and
\be
\Delta^{(n)}_{\text{\tt NMSSMCALC}} r = \Delta^{\text{lit.}}_{\sm} r + \Delta^{(n)}_{\susy} r
\,.
\label{eq:deltar2}
\ee
Note that $\Delta^{(n)}_{\sm} r$ should be computed in the same way as for $\Delta^{(n)}_{\nmssm} r$. 

The SM corrections from the literature consist of the following terms
\begin{align}
    \Delta_{\sm}^{\text{lit.}} r =& \Delta^{(1)} r + \Delta^{(\alpha \alpha_s)} r+ \Delta^{(\alpha \alpha_s^2)} r
+ \Delta^{(\alpha^2)} r \crn
&+  \Delta^{(G_\mu^2 m_t^4\alpha_s)} r+  \Delta^{(G_\mu^2 m_t^6)} r
+  \Delta^{(G_\mu^2 m_t^2\alpha_s^3)} r \,.
\label{eq:SMHO}
\end{align}
For a complete list of references we refer the reader to \cite{Heinemeyer:2013dia}. 
In the following, we refer to previous works that provide analytical/numerical results
which are used in \cref{eq:SMHO}.
The full two-loop QCD corrections  $\Delta^{(\alpha \alpha_s)} r$ are taken from \cite{Halzen:1990je}, 
partial three-loop QCD corrections $\Delta^{(\alpha \alpha_s^2)} r$ are taken
from \cite{Chetyrkin:1995js}.
The three-loop corrections $\Delta^{(G_\mu^2 m_t^4\alpha_s)} r$, $ \Delta^{(G_\mu^2 m_t^6)} r$
to the $\rho$ parameter are taken from \cite{Faisst:2003px} and the four-loop
QCD corrections $\Delta^{(G_\mu^2 m_t^2\alpha_s^3)} r$ from
\cite{Chetyrkin:2006bj,Boughezal:2006xk}. 
For the full two-loop electroweak corrections  $\Delta^{(\alpha \alpha_s^2)} r$ 
we use the fitting formula presented in \cite{Awramik:2006uz}.
For consistency, we need to use the running-width definition of the $W$ and $Z$ masses in the two-loop
electroweak corrections, \cf footnote 7 in \cite{Stal:2015zca}, while the
fixed-width masses are used in the rest of the $\Delta r$ calculation due to
the employed OS scheme.

\subsection{NMSSM-Specific Two-Loop Corrections}
\label{sec:deltarsusy2}
As discussed in the previous sections, corrections to $\Delta r$ can be
categorized into corrections arising from the renormalization of the electric
charge, vertex and box diagrams and other corrections arising from the $W$ 
and $Z$ boson self-energies. For the first three
categories, higher-order corrections 
in the (N)MSSM beyond the one-loop level are unknown and not considered in this
work. This should be a good approximation for most phenomenologically
viable scenarios since these corrections 
are expected to be numerically small.
The dominant contributions arise from the $W$ and $Z$ boson self-energies 
and can be parameterized in terms of the loop corrections to the $\rho$
parameter. Thereby, the two-loop corrections to the $\rho$ parameter,
$\Delta^{(2)}_{\susy} \rho$, discussed in \cref{sec:rho},  
can be computed in an efficient way. 
However, $\Delta^{(2)}_{\nmssm} \rho$ also includes the SM-like corrections which
are already taken into account in $\Delta^{\text{lit.}}_{\sm} r$. Hence,
the SM-like contributions have to be subtracted from the $\rho$ parameter in
order to avoid double counting.
We define the two-loop SUSY correction 
to $\Delta \rho$ as 
\be
\Delta^{(\alpha_i^2)}_{{\susy}} \rho = \Delta^{(\alpha_i^2)}_{\nmssm}
\rho -\Delta^{(\alpha_i^2)}_{\sm} \rho \,,
\ee 
with
$\alpha_i^2=\alpha_t\alpha_s,\,\left(\alpha_{\lambda}+\alpha_{\kappa}+\alpha_{t}\right)^2$. 
The final expression for $\Delta_{\susy}^{(2)} r$ including the one and two-loop
corrections, that is used together with \cref{eq:deltar2} in \eqref{eq:MWcalcit}
to compute the $W$ mass in \NMSSMCALC, reads
\be \Delta^{(2)}_{\susy} r = \Delta^{(1)}_{\susy} r - \fr{c_W^2}{s_W^2} (
\Delta_{\susy}^{(\alpha_t\alpha_s)}\rho
+\Delta_{\susy}^{\left(\alpha_{\lambda}+\alpha_{\kappa}+\alpha_{t}\right)^2}\rho
)\,,
\ee
where $\Delta^{(1)}_{\susy} r$ is defined via \cref{eq:deltar1}.

To cross-check the implementation of the known SM corrections as well as to
check the correct decoupling behaviour of the NMSSM-specific corrections, we
verified that our NMSSM $W$-mass prediction approaches the one in the SM, as found in
\cite{Heinemeyer:2013dia, Stal:2015zca}, once all NMSSM particle masses are
chosen to be well above the electroweak scale.
\section{Numerical Analysis}
\label{sec:analysis}
In this section, we investigate the phenomenological impact of the
$\orderew$ corrections on the electroweak $\rho$ parameter and on the $W$ boson
mass. Detailed studies of the one-loop and the leading two-loop
corrections can be found in \cite{Domingo:2011uf,Stal:2015zca,Athron:2022isz}.
Furthermore, we estimate the uncertainty due to missing higher-order
SUSY corrections to $\Delta\rho$ by changing the renormalization
scheme of the top/stop sector. Possible sources of uncertainties entering 
the $M_W$ prediction are discussed as well.
For illustrative purposes, the results are shown for a set of two
parameter points, one obtained from a simple scan,
which will be described in the next paragraph, and one taken from the literature.
For a full investigation of the viable parameter space where the NMSSM
can simultaneously explain the $(g-2)_\mu$ data, the Higgs data, the
$W$ mass, and where the lightest neutralino is still a good lightest
supersymmetric particle (LSP) and Dark Matter 
candidate, we refer the reader to the recent studies
\cite{Tang:2022pxh,Domingo:2022pde} for the NMSSM, and to
\cite{Bagnaschi:2022qhb} for the MSSM.
Finally, we compare the $W$-mass prediction in \NMSSMCALC with the one obtained
from the public NMSSM spectrum generators \FlexibleSUSY, \NMSSMTools and \SASP.

\subsection{Setup of the Parameter Scan \label{subsec:scan}}
In order to find parameter points which are not excluded by the
measured properties of the \unit[125]{GeV} Higgs boson and by the
direct LHC searches for SUSY particles, we have performed a scan over
the NMSSM parameter space in the following way. For a given set of
input parameters, we use {\tt NMSSMCALC} to calculate the mass spectrum of
all SUSY particles at the one-loop order. 
In addition, the Higgs boson masses are obtained including the available two-loop
corrections at ${\cal O}(\al_s\al_t +(\al_t+\al_\lambda
+\al_\kappa)^2)$ \cite{Dao:2021khm}. The Higgs boson masses and mixing
matrices are computed using OS renormalization in the top/stop
sector. One of the neutral CP-even Higgs bosons is identified with the
SM-like Higgs boson, and its mass is required to lie in the range 
\beq
\unit[122]{GeV} \le m_h \le \unit[128]{GeV} \;.
\eeq  
In the following, $m_h$ will always denote the mass of the SM-like Higgs
boson which not necessarily needs to be the mass $m_{h_1}$ of the lightest
scalar state.
We further require that 1) the lightest SUSY particle is the lightest
neutralino; 2) the masses of the electroweakinos and sleptons satisfy
the lower bounds from LEP \cite{ParticleDataGroup:2022pth}; 3) the
masses of the stops, lightest sbottom and the gluino
satisfy the lower bounds taken from the 2022 Particle Data 
Group review \cite{ParticleDataGroup:2022pth}, in particular
\begin{align}  M_{\chi_1^\pm} & > 94\, \gev, &  M_{\chi^0_2}& > 62.4\, \gev, & M_{\chi^0_3}& > 99.9\, \gev,    \crn
M_{\chi^0_4}& > 116\, \gev, & M_{\ti e_1}& > 107\, \gev, & M_{\ti \mu_1}& > 94\, \gev, \crn
M_{\ti \tau_1}& > 81.9 \, \gev, & M_{\ti \nu_{e/\mu/\tau}}& > 94 \, \gev, & M_{\ti b_1}& > 1270 \, \gev, \crn
M_{\ti t_1}& > 1310 \, \gev, & M_{\ti g}& > 2300\, \gev.  \end{align}
The Higgs decay widths and branching
ratios including the state-of-the-art higher-order QCD corrections
as well as the effective Higgs couplings, {\it i.e.} using the Higgs
mixing angles obtained from the diagonalization of the loop corrected
mass matrices, are obtained with \NMSSMCALC, too.
Having all important properties of the Higgs sector at hand, we use {\tt
HiggsTools} \cite{Bahl:2022igd} which contains {\tt HiggsBounds-5}
\cite{Bechtle:2020pkv}, to check if the parameter points pass all the exclusion
limits from the searches at LEP, Tevatron and the LHC, and {\tt HiggsSignals-2}
\cite{Bechtle:2020uwn} to check if the points are consistent with the LHC data
for a \unit[125]{GeV} Higgs boson within 2$\sigma$.
To constrain the SUSY fermionic and scalar sector we use {\tt SModelS-2}
\cite{Alguero:2021dig} to check whether a given scenario is excluded
by the LHC searches for the electroweakinos and
sleptons. For the input of {\tt SModelS-2}, we have implemented the chargino and
neutralino decay widths and branching ratios in a private version of
\NMSSMCALC.
Furthermore, the production cross sections of all pairs of
electroweakinos have been computed at leading order 
using a private implementation that was obtained with the help of {\tt
  FeynArts-3.11} \cite{Hahn:2000kx} and 
{\tt FormCalc-9.8} \cite{Hahn:formcalc}. These tree-level cross sections are
corrected by a common $K$-factor to account for the NLO QCD corrections.
We chose $K=1.3$ which was obtained with the help of {\tt Prospino2.0}
\cite{Beenakker:1996ed,Beenakker:1999xh}. 

Since there are many experimental constraints applied in the scan, using a 
uniform random scan over all input parameters can be very time and
resource consuming. We therefore performed a Markov Chain Monte Carlo
sampling using {\tt EasyScan\_HEP-1.0} \cite{Shang:2023gfy} in order to efficiently find
phenomenologically viable parameter regions. Table~\ref{tab:nmssmscan}
summarizes the ranges applied in the parameter scan.
\begin{table}
\begin{center}
\begin{tabular}{l|ccc|cccccccc} \toprule
& $t_\beta$ & $\lambda$ & $\kappa$ & $M_1,M_2$  & $A_t$ &
 $m_{\tilde{Q}_3}$ & $m_{\tilde{t}_R}$ & $m_{\tilde{e}_{iR}},m_{\tilde{L}_{i}}$ & $M_{H^\pm}$
& $A_\kappa$ & $\abs{\mu_{\text{eff}}}$ \\
& & &  \multicolumn{9}{c}{in TeV} \\ \midrule
min &  1 & 0 & -1 & 0.1  & -5 &  1 & 1 & 0.1 & 0.6 &-3 & 0.1 \\
max & 20 & 1 & 1 & 1    & 5  & 3   &  3 & 2   & 3   & 3 & 2 \\ \bottomrule
\end{tabular}
\caption{Input parameters for the NMSSM scan. All parameters have been 
varied independently between the given minimum and maximum
values. \label{tab:nmssmscan}}
\end{center}
\vspace*{-0.6cm}
\end{table}
The SM input parameters, that are relevant for the calculation of $\Delta\rho$ and
the  $W$ mass\footnote{In the calculation of all Higgs masses
$\alpha(M_Z) = 1/127.955$ and $M_W= 80.377~\gev$ are chosen as input
parameters.}, are taken from \cite{ParticleDataGroup:2022pth}
\begin{equation}
\begin{tabular}{lcllcl}
\quad $\alpha_0$ &=& 1/137.035999084\,, &\quad $\alpha^{\overline{\mbox{MS}}}_s(M_Z)$ &=&
0.1179\,, \\
\quad $G_F$ &=& $1.1663788 \cdot 10^{-5}$~GeV$^{-2}$\,, &\quad   $M_Z$ &=& 91.1876~GeV \,, \\
\quad $m_t$ &=& 172.69~GeV\,, &\quad $m^{\overline{\mbox{MS}}}_b(m_b^{\overline{\mbox{MS}}})$ &=& 4.18~GeV\,, \\
\quad $m_\tau$ &=& 1.77686~GeV\,, &\quad $\Delta\alpha_{\text{had}}^{(5)}(M_Z^2)$&=& 0.02768\,,\\
\quad$\Delta\alpha_{\text{lepton}}(M_Z^2)$ &=& 0.03150\,.&\quad  
\end{tabular}
\end{equation}
For the following numerical analysis, 
we have chosen two parameter points to study the impact of the new corrections. The
first parameter point, called {\tt P1}, which passes all our constraints specified
above, is given by the following input parameters,
\beq
&\text{\tt P1}:\quad&  
m_{\tilde{t}_R}=2002\,\gev \,,\;
 m_{\tilde{Q}_3}=2803\,\gev\,,\; m_{\tilde{b}_R}=2765\,\gev\,, \; \non\\ 
 && 
m_{\tilde{L}_{1,2}}= 565\,\gev\,,\; m_{\tilde{e}_R,\tilde{\mu}_R}=374\,\gev \,,\; 
 \non\\
&& 
m_{\tilde{L}_{3}}= 575\,\gev\,,\; m_{\tilde{\tau}_R}=981\,\gev \,,\; 
 \non\\ 
&& |A_{u,c,t}| = 2532\,\gev\, ,\; |A_{d,s,b}|=1885\,\gev\,,\;
|A_{e,\mu,\tau}| =1170\,\gev\,,\; \label{eq:paramP1} \\ \non
&& |M_1| = 133\,\gev,\; |M_2|= 166\,\gev\,,\; |M_3|=2300\,\gev \,,\\ \non
&& \lambda = 0.301\,\gev,\; \kappa= 0.299\,\gev\,,\; \tan\beta=4.42\,\gev \,,\\ \non
&& \mu_{\text{eff}} = 254\,\gev,\; \text{Re}A_\kappa= -791\,\gev\,,\;  M_{H^\pm}= 1090\,\gev\,, \\ \non
&&  \varphi_{A_{e,\mu,\tau}}=0\, ,\; \varphi_{A_{d,s,b}}=\pi\,,\; 
\varphi_{A_{u,c,t}}=\varphi_{M_1}=\varphi_{M_2}=\varphi_{M_3}=0
 \;.
\eeq
The resulting spectrum of the Higgs boson masses at  ${\cal
  O}(\al_s\al_t +(\al_t+\al_\lambda 
+\al_\kappa)^2)$  is given in \tab{tab:HiggsSpectrumP1OS}. The SM-like
Higgs boson is dominated by the $h_u$ component and its mass 
is about 125.4 GeV, the remaining Higgs bosons are heavier. The spectrum of the
electroweakinos  is given in \tab{tab:weakinoP1OS}. 
For this parameter point, the lightest neutralino is the lightest SUSY particle.
While the masses of the electroweakinos are rather
  light with masses below about 510~GeV, the masses of sleptons are rather
heavy, larger than $\unit[380]{GeV}$.  The lightest
sleptons are mainly composed of the right-handed selectron and smuon components.
The special feature of this point is that the wino, bino and
higgsino components mix strongly. Thus, one can not distinguish which neutralino
is wino-, bino- or higgsino-like, which is a region where experimental
constraints are rather weak. In particular, the cross sections of the 
electroweakino pair production processes 
become small and therefore the parameter point escapes the LHC constraints from
the electroweakino searches.

\begin{table}[t!]
\begin{center}
\begin{tabular}{|c|c|c|c|c|c|c|}
\hline
                           ${H_1}$   
                              &  ${H_2}$  &  ${H_3}$  &  ${H_4}$  
                              &  ${H_5}$ 
                               \\ \hline \hline
125.4 & 230.6 & 770.5  & 1088.0 & 1090.1
              \\ \hline
  $ h_u$     & $h_s$   & $a_s$   & $a$   & $h_d$
  \\ \hline
\end{tabular}
\caption{The parameter
  point {\tt P1}:  Higgs  masses in GeV and  the main components are shown.}
\label{tab:HiggsSpectrumP1OS}
\end{center}
\vspace*{-0.4cm}
\end{table} 
\begin{table}[t]
\begin{center}
\begin{tabular}{|c|c|c|c|c|c|c|c|c|}
\hline
                           ${\ti \chi_1^0}$   
                              & ${\ti \chi_2^0}$ & ${\ti \chi_3^0}$ & ${\ti \chi_4^0}$ 
                              & ${\ti \chi_5^0}$
                              & ${\ti \chi_1^+}$& ${\ti \chi_2^+}$ \\ \hline \hline
113.9 & 145.07 & 261.95 & 295.74   & 509.49& 132.93 & 294.98  
              \\ \hline
\end{tabular}
\caption{Electroweakino  masses in GeV for the parameter
  point {\tt P1}.}
\label{tab:weakinoP1OS}
\end{center}
\vspace*{-0.4cm}
\end{table} 

\subsection{Results for the $\rho$ Parameter}
\begin{figure}[t]
    \centering
    \includegraphics[width=0.485\textwidth]{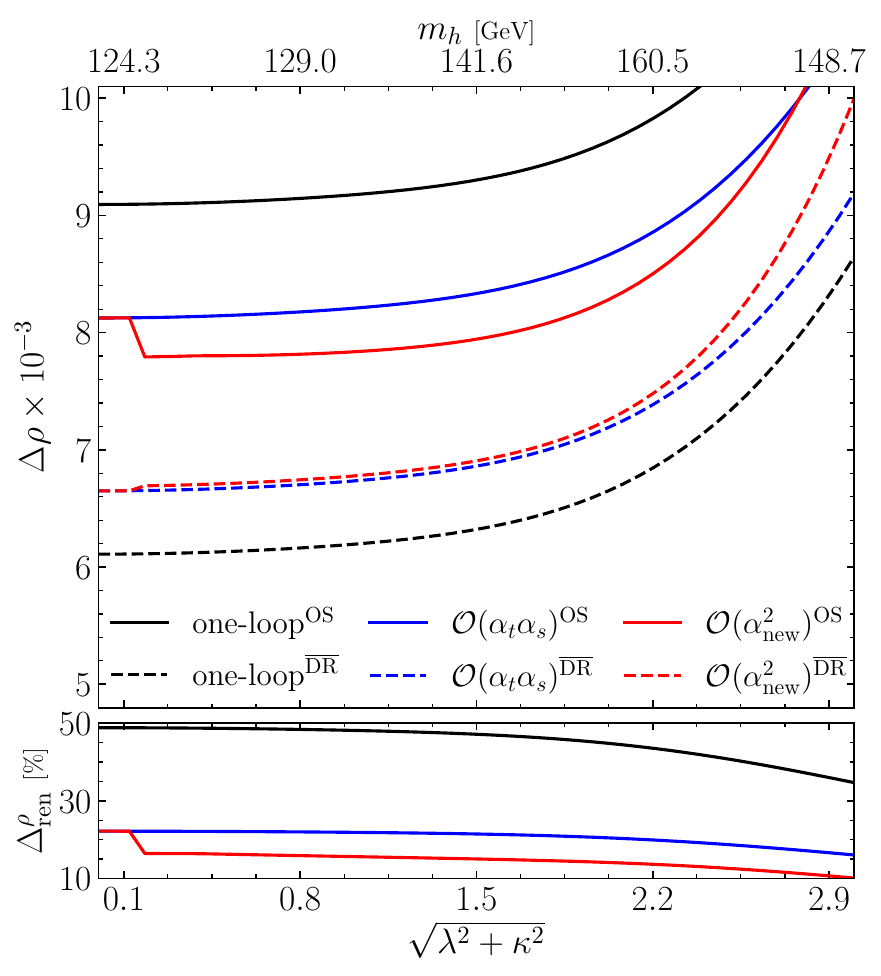}
    \includegraphics[width=0.5\textwidth]{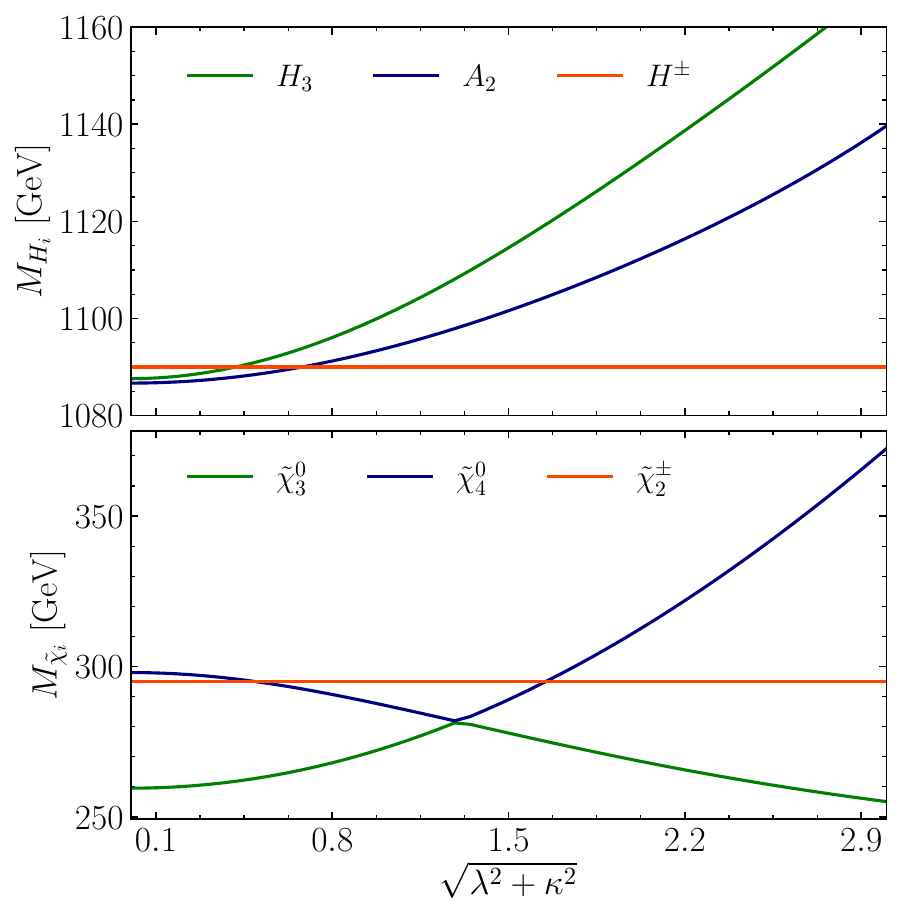}
    \caption{Upper left: The $\rho$ parameter at full
      one-loop order (black),
      two-loop $\orderqcd$
      (blue) and two-loop $\ordernew$ (red) as function  of
      $\sqrt{\lambda^2 + \kappa^2}$. 
    The dashed lines are for the $\DRbar$ scheme in the top/stop
    sector while the full lines are for the OS scheme. Lower left:
    Renormalization scheme dependence $\Delta^\rho_{\text{ren}}$ of
    the $\rho$ parameter at full one-loop order (black),
    at two-loop $\orderqcd$ (blue) and $\ordernew$ (red). 
    Right: The part of the Higgs- and EW-ino mass spectrum causing
    larger deviations.}
    \label{fig:rho}
\end{figure}

\newcommand{\slk}{s_{\lambda\kappa}}
In the following, we investigate the prediction for the $\rho$
parameter starting from the
parameter point {\tt P1} as a function of the NMSSM-specific superpotential parameters
$\lambda$ and $\kappa$. In order to avoid negative squared tree-level masses, we
simultaneously vary both $\lambda$
and $\kappa$ around their values $\lambda_0=0.301$ and
$\kappa_0=0.299$ such that the ratios $\lambda/\lambda_0$ and
$\kappa/\kappa_0$, respectively, are kept equal. Furthermore, we vary
$A_\kappa=A_{\kappa_0}\left(1-(\lambda-\lambda_0)/4\right)$ to avoid a tachyonic
tree-level Higgs mass even for very large values of $\lambda$. All other parameters of {\tt P1} are kept fixed.
The upper panel of \cref{fig:rho} (left) shows the $\Delta\rho$ parameter computed at
one-loop (black), $\orderqcd$ (blue) and $\ordernew$ (red) as a
function of $\sqrt{\lambda^2+\kappa^2}$ (lower axis) and the obtained Higgs mass
including all available two-loop corrections (upper axis). We here
introduced for better readability the notation
\begin{subequations}
\begin{align}
    \alpha_{\text{new}}^2 &\equiv \alpha_t \alpha_s + (\alpha_\lambda +
\alpha_\kappa + \alpha_t)^2 \;,\\
    \slk &\equiv\sqrt{\lambda^2+\kappa^2}\;.
\end{align}
\end{subequations}
Since the prediction of the $\rho$ parameter strongly
depends on the top quark mass, it is shown using both the OS (full
lines) and the $\DRbar$ (dashed lines) renormalization schemes in the 
top/stop sector. The $W$ boson mass entering the $\rho$ parameter
prediction has been chosen to be the value for $M_W$ obtained at
$\ordernew$, \cf \cref{sec:MWp1}. 

Note that points with $\slk> 0.7$ generally are in danger to
violate perturbative unitarity and/or to run into a Landau pole close to their
input scale. Nonetheless, it is possible to go to larger values in some regions
of the parameter space (giving up the requirement that the NMSSM is UV-complete).
From this point of view, it is interesting to study which values of
$\lambda,\kappa$ would be required in order to obtain sizeable
corrections. We therefore plot our results for values up to
$\slk=3$.
In the OS scheme, the $\orderqcd$ and $\ordernew$ corrections are
both negative.
In the region that is free of low-energy Landau poles, the $\orderqcd$ corrections reduce the one-loop result by 10\% while the $\ordernew$ corrections reduce the $\orderqcd$ results by 4\%.
In contrast, the $\orderqcd$ and  $\ordernew$ corrections are both positive in the $\DRbar$ scheme.
The $\orderqcd$ corrections increase the one-loop value by 9.8\% while the
$\ordernew$ corrections add 0.6\% on top of the $\orderqcd$ results. In the 
region  $\slk>2$, the $\orderew$ corrections in the
two renormalization schemes start to deviate 
differently from the $\orderqcd$ 
corrections. The magnitude of the $\orderew$ corrections becomes smaller in the OS scheme, but larger in the $\DRbar$ scheme. 

We observe an increase in $\Delta\rho$ with increasing $\slk$
which first is rather weak and becomes
stronger for very large values of $\slk$. This behaviour is correlated
with an increase of the  $SU(2)$ mass splittings
between the neutral and charged Higgs bosons on the one side and the
neutral and charged electroweakinos on the other side. This can be inferred from the right plots, where we
show in the upper plot the dependence of the
charged Higgs mass and the heavy CP-even/odd Higgs masses, which have a
dominant $h_d$ and $a_d$ component, respectively, as a function of
$\slk$, while the masses of their 
corresponding EW-ino states as function of $\slk$ are shown in the lower plot.

We conclude this section with a discussion about numerical instabilities that can appear
in the $\orderew$ corrections to $\Delta \rho$, especially for small values of $\lambda$ and $\kappa$.
The corrections to the $\Delta\rho$ parameter are composed of $W$ and $Z$
boson self-energies which individually can receive very large higher-order corrections. However, in
the difference of the self-energies, entering $\Delta\rho$, large cancellations
of many orders of magnitude can appear. In some cases the size of this
cancellation may exceed the numerical precision of the program which is
currently limited to {\tt double} precision. The $\orderew$
corrections are particularly sensitive to this kind of instabilities since the
tree-level masses, which enter the two-loop self-energies, are calculated at
$\mathcal{O}(\alpha_\lambda+\alpha_\kappa)$ in the gaugeless limit which can
lead to very small but non-zero tree-level masses that enter the calculation of
the two-loop self-energies.
While these instabilities might not be apparent in the $\rho$ parameter
prediction in the first place, the $M_W$ prediction which discussed in
\cref{sec:MWp1}, is very sensitive to the value of $\Delta \rho$. In particular
the VEV (and all parameters and couplings derived from it) depends on the floating value
of $M_W$ during the iteration which can easily amplify numerical instabilities
coming from $\Delta \rho$. Therefore, the $M_W$ prediction in  {\tt NMSSMCALC} at $\orderew$ can only be used
for parameter points that do not suffer from large numerical instabilities. If
no convergence is found at $\orderew$ in $M_W$, the program
automatically falls back to 
the $\orderqcd$ predictions for both the $W$-mass and $\Delta \rho$.
Furthermore, some parameter points may feature tachyonic tree-level masses in the gaugeless limit, while the actual tree-level masses are positive. While Ref.~\cite{Domingo:2021jdg} proposed a
possible solution to a similar problem at the one-loop order, our calculation is mostly
affected by tachyonic states required for partial two-loop corrections.
The generalisation of the method developed in Ref.~\cite{Domingo:2021jdg} is
beyond the scope of this work and left for future work. Therefore, if the program
encounters negative squared tree-level masses in the preparation of the masses
entering at a given loop-order, it also automatically falls-back to the
next-lowest order that is not involving tachyonic tree-level masses in the
Feynman diagrams\footnote{If the full tree-level masses, which enter the
    full one-loop mass prediction, are found to be tachyonic, the program exits
with an error message.}.

In \cref{fig:rho} we observe good convergence for $\slk>0.1$ for $M_W$ at
$\ordernew$, \cf \cref{sec:MWp1}. For $\slk<0.1$ the $M_W$ prediction does not
converge and therefore the $\rho$ parameter prediction is used at $\orderqcd$
which explains the jump of the red line onto the blue line.\s
{\flushleft\underline{\textbf{Uncertainty Estimate for $\Delta\rho$}}}\\
In order to estimate the uncertainty due to missing higher-order corrections to
the $\rho$ parameter, we define the renormalization scheme dependence of
the $\rho$ parameter at a given loop order as 
\beq
\Delta_{\text{ren}}^{\rho}=\fr{\Delta\rho^{\text{OS}}-\Delta\rho^{\DRbar}}{\Delta\rho^{{\DRbar}}}.
\eeq
The lower panel of \cref{fig:rho} (left) shows the resulting
$\Delta_{\text{ren}}^{\rho}$ obtained at the three considered loop
orders as a function of $\slk$.
We observe a renormalization scheme dependence of up to $50\%, 22\%$ and $16\% $ for the one-loop, $\orderqcd$ and  $\ordernew$ results, 
respectively. Therefore, including the two-loop QCD and EW corrections can
significantly reduce the theory uncertainty of the $\rho$ parameter.
For the comparison with the SM we present in \cref{tab:SMrho} the $\rho$
parameter computed in the SM at the corresponding one-loop order and
  two-loop ${\cal O}(\alpha_t \alpha_s)$ and ${\cal O}(\alpha_t
  \alpha_s + \alpha_t^2)$
   using $\DRbar$ and OS renormalisation
conditions in the top/stop sector. In the SM, the $\orderqcd$ and ${\cal
  O}(\alpha_t\alpha_s+ \alpha_t^2)$ corrections are negative both in
the OS and the $\DRbar$ scheme. 
The renormalisation scheme dependence $\Delta_{\text{ren}}^{\rho}$ in
the SM is $55\%,\, 45\%$ 
and $42\%$ at one-loop order, $\orderqcd$ and ${\cal O}(\alpha_t\alpha_s+
\alpha_t^2)$, respectively, which is significantly larger than the
corresponding results obtained in the NMSSM. Therefore, the SUSY QCD
contributions to the $\rho$ parameter seem to play an important role
in the reduction of the scheme dependence of the $\rho$ parameter.
We want to stress that it is not possible to draw conclusions about the scheme
uncertainty of the $M_W$ prediction from the scheme uncertainty of the $\rho$
parameter. In particular for the SM prediction of $M_W$ a full two-loop
prediction, beyond corrections to $\Delta \rho$, was found to yield only an uncertainty of $\order{\unit[3-7]{MeV}}$
\cite{Degrassi:2014sxa} implying cancellations of the scheme dependence between
$\Delta\rho$ and other quantities entering $\Delta
r$. For a discussion of the $M_W$ uncertainty in the
  NMSSM, see the next section. 
\begin{table}[t]
\begin{center}
\begin{tabular}{|c|c|c|c|}
\hline
& one-loop & $\orderqcd$ & ${\cal O}(\alpha_t\alpha_s+ \alpha_t^2)$ \\ \hline \hline
$\DRbar$ & 5.734 & 5.478 & 5.295     
              \\ \hline
OS & 8.918 & 7.950 & 7.509 \\ \hhline{|=|=|=|=|}
$\Delta_{\text{ren}}^{\rho}$ [\%] & 55 & 45 & 42 \\ \hline
\end{tabular}
\caption{The SM $\rho$ parameter multiplied by a factor of $10^{-3}$, 
computed with $M_W=80.36\,\gev$ and $m_h=125\,\gev$ at one-loop and
two-loop ${\cal O}(\alpha_t \alpha_s)$ and ${\cal O}(\alpha_t \alpha_s
+ \alpha_t^2)$.}
\label{tab:SMrho}
\end{center}
\end{table} 

\subsection{Results for the $W$ Boson Mass}
\label{sec:MWp1}
In this section we discuss the prediction of $M_W$ at one-loop order, two-loop 
$\orderqcd$ and $\ordernew$.  It is important to stress that a consistent
inclusion of the known higher-order SM corrections to $\Delta r$
beyond two-loop order, \cf \cref{sec:smho}, is only 
possible if the SM sector entering the SUSY corrections is renormalized in the
same renormalization scheme as in the SM calculation that has been implemented
in \NMSSMCALC. Therefore, we exclusively
chose the OS scheme in the top/stop sector entering the prediction of
$M_W$.

We define two quantities to investigate the behaviour of $M_W$:
\newcommand{\deltasm}{\Delta^{\text{SM}}_{m_h}}
\begin{subequations}
\begin{align}
    \deltasm &= M_W^{\text{NMSSM}} - M_W^{\text{SM}}(m_h) \label{eq:deltasm}
    \quad\\ 
  \text{and}\quad   \Delta^{\alpha_i}_{\alpha_j} &=
        M_W^{(\alpha_i)}-M_W^{(\alpha_j)}.\label{eq:deltas}
\end{align}
        
\end{subequations}
The quantity $\deltasm$ defines the difference between the NMSSM $W$-mass
prediction at a given order and the SM prediction, including all SM higher-order
corrections, evaluated with the Higgs boson pole-mass $m_h$ that is predicted
by the NMSSM for the considered scenario at the two-loop level. Therefore, also 
$M_W^{\text{SM}}(m_h)$ varies with $\lambda$ and $\kappa$ as the Higgs
mass prediction in the NMSSM changes. Since the SM
higher-order corrections drop out in $\deltasm$, it can be used as a measure
for the size of the genuine SUSY corrections to $M_W$. The quantity
$\Delta^{\alpha_i}_{\alpha_j}$ determines the size of specific
higher-order SUSY correction $\alpha_i$ w.r.t. the next-lowest order $\alpha_j$.

\begin{figure}[t]
    \centering
        \includegraphics[width=0.53\textwidth]{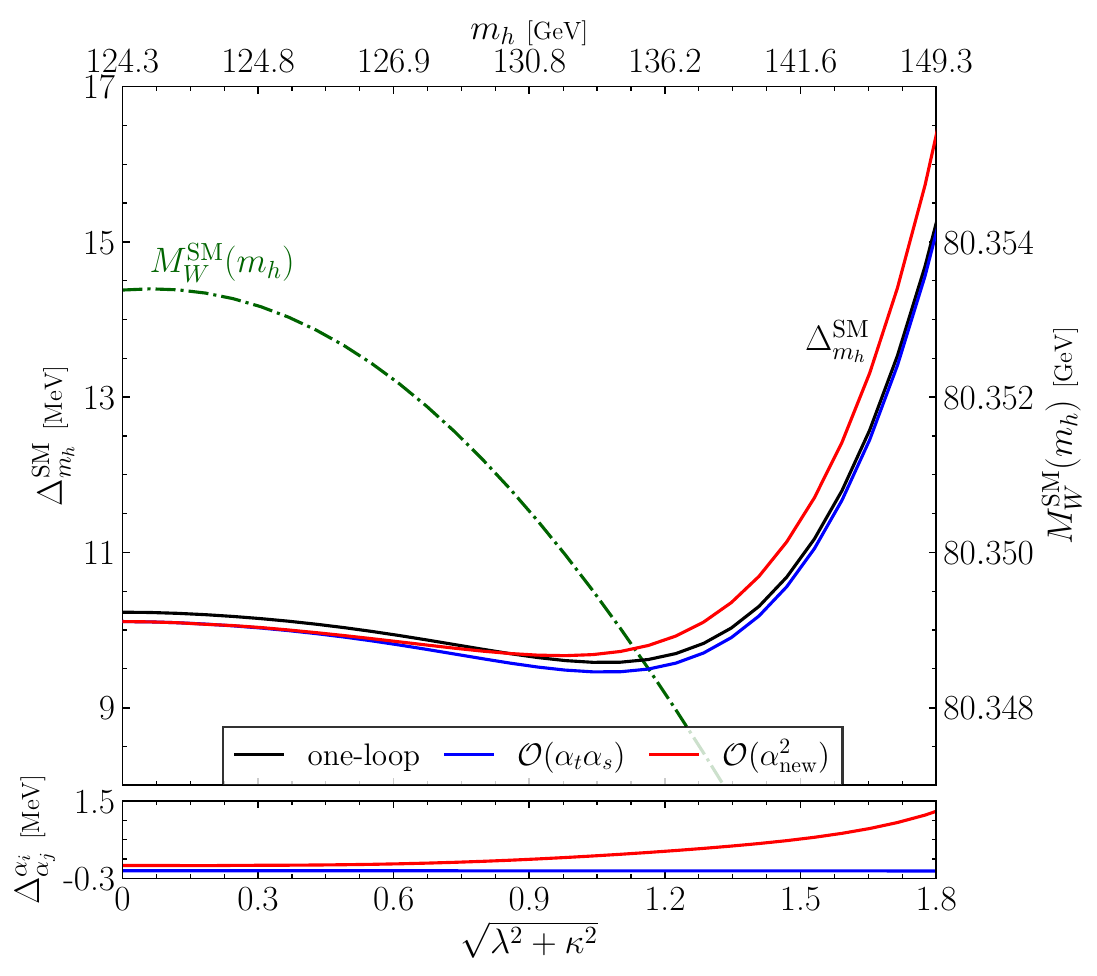}
        \includegraphics[width=0.46\textwidth]{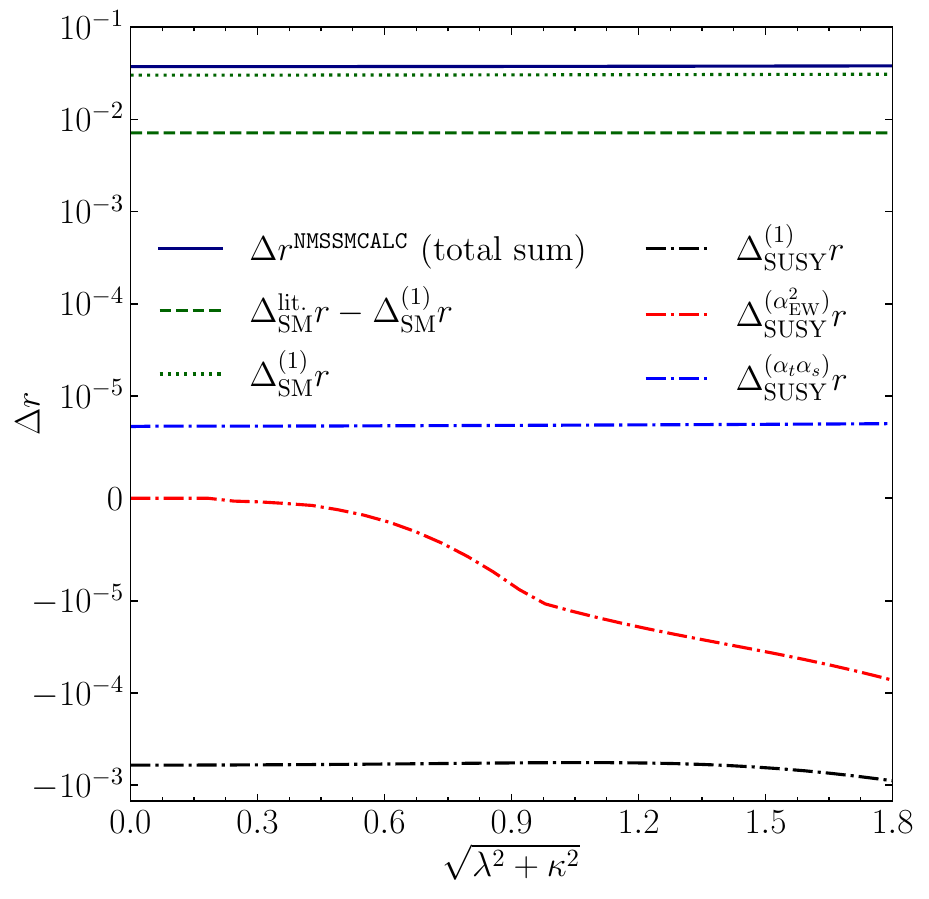}
    \caption{Upper left: the difference $\deltasm$ (left axis) defined in \cref{eq:deltasm}   at one-loop (black), two-loop $\orderqcd$ 
      (blue) and two-loop $\ordernew$ (red) as a
      function of $\sqrt{\lambda^2 + \kappa^2}$, and $M_W$ in the SM (green dash-dotted, right axis) using the
      loop-corrected NMSSM Higgs boson mass prediction (upper axis).
      Lower left: the difference $\Delta^{\alpha_t\alpha_s}_{\text{one-loop}}$ (blue) and
      $\Delta^{\alpha_{\text{new}}^2}_{\alpha_t\alpha_s}$ (red) defined via
      \cref{eq:deltas}.
    Right: Individual contributions to $\Delta r$  predicted by \NMSSMCALC. For $|\Delta r|>10^{-5}$ a log-scale is
            chosen and a linear scale otherwise.} 
    \label{fig:MW}
\end{figure}

Figure~\ref{fig:MW} (upper left) shows $\deltasm$ (left axis) as a
function of $\slk=\sqrt{\lambda^2+\kappa^2}$ 
starting from the parameter point {\tt P1} at one-loop (black solid) and two-loop $\orderqcd$
 (blue solid) and $\ordernew$ (red solid). Note that we have used 
 the same procedure for the variation of the parameters described in the previous section. The upper axis shows the obtained
 value for $m_h$, including all available two-loop corrections, that is used in
the prediction of $M_W^{\text{SM}}(m_h)$ (green dot-dashed, right axis). The lower panel shows
$\Delta^{\alpha_t\alpha_s}_{\text{one-loop}}$ (blue) and
$\Delta^{\alpha_{\text{new}}^2}_{\alpha_t\alpha_s}$ (red).
We observe that the NMSSM-specific corrections range between about 
\unit[10-20]{MeV},
mostly dominated by the one-loop corrections. The QCD corrections to the $W$ boson mass are negative
compared to the one-loop prediction and subtract about
  0.2~MeV from the one-loop result, independent of the value of
  $\slk$. Compared to the ${\cal O}(\alpha_t \alpha_s)$ result the $\ordernew$ corrections
range between \unit[-0.2]{MeV} and \unit[+1.5]{MeV} in the shown range
of $\slk$.

To get a better understanding of the individual contributions to the $W$-mass
prediction we plot the values of $\Delta r$ obtained after the $M_W$ iteration
has converged. In \cref{fig:MW} (right) the blue solid line shows the total
result of $\Delta r$ obtained with \NMSSMCALC including all available
corrections. The green dotted line shows the one-loop SM-contributions which also
contain the $\Delta\alpha$ contributions that are numerically most significant.
The green dashed line shows the size of the higher-order SM results taken
from the literature which are the next-to-largest contribution to $\Delta r$.
The third-largest contribution is the one-loop SUSY contribution (black
dash-dotted) which is negative (hence a positive shift to $M_W$) followed by the
-- also negative -- EW contributions (red dash-dotted). The SUSY QCD corrections
are positive and numerically in competition with the EW corrections for
$\slk\gtrsim 0.6-0.9$. Note that in \cref{fig:MW} (right) we chose a log-scale
for $|\Delta r|>10^{-5}$ and a linear scale otherwise which sets the focus on
the $\slk$-dependence of the two-loop SUSY corrections.

For the other parameter points which pass the constraints of our scan
described in \cref{subsec:scan} we observe similar
features. The $\orderqcd$ and $\ordernew$
corrections to $M_W$ are rather small and their sign w.r.t.~to the one-loop
corrections can change, depending on the 
considered parameter point.\s

{\flushleft\underline{\textbf{Uncertainty Estimate for $M_W$}}}\\
\label{sec:MWuncert}
In this paragraph we comment on the uncertainties that contribute to the $M_W$ calculation in
{\tt NMSSMCALC}. We focus on two sources of uncertainties: 
(i) parametric uncertainties that are introduced through the dependence on
experimental input parameters and (ii) theory uncertainties due to the
missing higher-order corrections. To estimate (i) we vary the SM input
parameters $M_{Z}$, 
$\Delta\alpha_{\text{had}}^{(5)}(M_Z^2)$, $\alpha_s $ within their $1\sigma$ PDG
values \cite{ParticleDataGroup:2022pth}. The top quark mass is varied by
\unit[1]{GeV}. \cref{tab:P1BP3uncer} lists the maximal differences in the  $W$
mass prediction compared to the result obtained at the central values for the 
parameter points {\tt P1}, {\tt BP3} (which is introduced in \cref{sec:comp})
and the SM prediction that includes all known higher-order corrections.
Furthermore, we also vary the result of the loop-corrected Higgs boson mass by
\unit[1]{GeV} to account for the theoretical uncertainty in the Higgs mass
prediction which is indirectly influencing the prediction of the $W$ boson mass.
We observe that the SUSY prediction does not introduce a significantly larger
parametric uncertainty compared to the SM prediction. The values for the latter
are in good agreement with those found in
\cite{Bagnaschi:2022qhb,Athron:2022isz}.
\begin{table}[t]
\begin{center}
\begin{tabular}{|c|c|c|c|c|}
\hline
& \multicolumn{3}{c|}{$\Delta M_W$ in \unit[MeV]} \\ \hline
                        & {\tt P1} & {\tt BP3} & {\tt SM}  \\ \hline \hline
$M_{Z}\pm 1 \sigma$     & 1.5      & 1.5       & 1.5       \\ \hline
$m_t\pm\unit[1]{GeV}$   & 5.6      & 5.8       & 5.6       \\ \hline
$\Delta\alpha_{\text{had}}^{(5)}(M_Z^2) \pm 1\sigma$ 
                        & 1.27     & 1.28      & 1.27     \\ \hline 
$\alpha_s \pm 1\sigma $ & 0.5      & 0.6       & 0.5      \\ \hline                           
$m_h \pm\unit[1]{GeV} $ & 0.47     & 0.47      & 0.47      \\ \hline 
\end{tabular}
\caption{The variation in the $W$ mass prediction around the central value 
when varying $M_{Z}$, $\Delta\alpha_{\text{had}}^{(5)}(M_Z^2)$, $\alpha_s$
within $1\sigma$ of their central values and $m_t$ within
$\unit[172.69\pm 1]{GeV}$, for {\tt P1} and {\tt BP3}. The Higgs mass $m_h$ used
in the SM-prediction is chosen to be the loop-corrected Higgs mass of the
respective parameter point and varied by $\pm\unit[1]{GeV}$.}
\label{tab:P1BP3uncer}
\end{center}
\end{table}
Combining all uncertainties in \cref{tab:P1BP3uncer} quadratically yields a total
parametric uncertainty of about \unit[6]{MeV}.
To estimate the size of the missing of
higher-order corrections, we first divide them into 
SM-like and SUSY corrections and discuss their individual 
uncertainties. Uncertainty estimates for the SM corrections have been studied in 
\cite{Awramik:2003rn,Degrassi:2014sxa,Bagnaschi:2022qhb} and yield about 4 MeV
and 3 MeV in the OS and $\MSbar$ calculation, respectively. A comparison between OS
and $\MSbar$ result, however, suggest an uncertainty of about \unit[6]{MeV}
\cite{Degrassi:2014sxa}.
Regarding the missing higher-order SUSY corrections, one can expect them to not  
be significantly larger than the computed $\orderqcd$ and $\ordernew$
corrections in large parts of the parameter space.
For the parameter points of our scan which pass the applied
constraints, the maximal corrections for $M_W$ are about \unit[4]{MeV} and
\unit[2]{MeV}, respectively. However, there remains the possibility that two-loop
SUSY corrections proportional to the electroweak gauge couplings, which are
unknown so far, could be enhanced in cases where $SU(2)$ states have large mass splittings.
Therefore, the SUSY uncertainty should be at least about \unit[4]{MeV} large. 

\subsection{Comparison with Previous $M_W$ Results}
\label{sec:comp}
Among publicly available tools, not only \NMSSMCALC is able to predict the
$W$-mass in the NMSSM at high precision.
In light of the CDF measurement, the spectrum-generator
generators \FlexibleSUSY \cite{Athron:2014yba,Athron:2017fvs} and {\tt SARAH}
\cite{Staub:2009bi,Staub:2010jh,Staub:2012pb,Staub:2013tta} have been updated to
be able to predict $M_W$ in a wide class of BSM models.
So far, the updated implementation in \SASP has been used to study $M_W$ in Dirac
gaugino models \cite{Benakli:2022gjn} while the one of \FlexibleSUSY was applied
to the MSSM, MRSSM and a singlet extended SM \cite{Athron:2022isz}. However,
they are in principle also capable to generate spectrum-generators that allow to
study $M_W$ in the NMSSM.
Furthermore, the program \NMSSMTools was extended in \cite{Domingo:2011uf} to
compute the $W$-mass in the general NMSSM as well as the
$\mathbb{Z}_3$-symmetric NMSSM described by \cref{eq:super}.
In the following, we briefly review the main ingredients for the $M_W$ prediction implemented in
\FlexibleSUSY, \NMSSMTools and \SASP while focusing on treatments that are
different from the \NMSSMCALC implementation described in \cref{sec:WMass}.
Following the discussion of the four different $M_W$ calculations in the NMSSM, we
numerically compare the prediction obtained for two concrete benchmark points.
Higher-order corrections to the muon anomalous magnetic moment $a_{\mu}$ are
known to have a connection to large corrections to $M_W$ \cite{Bagnaschi:2022qhb}. Since $a_\mu$ is
of increasing recent theoretical and experimental interest
\cite{Muong-2:2021ojo,Athron:2021evk,Dao:2022rui,Tang:2022pxh}, we also include a
comparison of the $a_\mu$ prediction between the various codes.

We start by discussing the incorporation of the SM higher-order corrections in
the different codes. \NMSSMCALC implements
the results of $\Delta_{\text{SM}}^{\text{lit.}} r$ to a large degree
analytically (\cf \cref{sec:smho}), while the other
three codes are based on fit formulas for the $M_W$ prediction within the SM.
In contrast to \NMSSMCALC, 
\FlexibleSUSY and \SASP compute
all BSM corrections to $M_W$ in the $\MSbar$/$\DRbar$ scheme and therefore rely on the SM
$\MSbar$ fit formula for $M_W^{\text{SM}}$ provided in \cite{Degrassi:2014sxa}:
\begin{equation}
    M_W^{{\substack{\FlexibleSUSY\\ \SASP}}} =
       \sqrt{ M_W^{\text{SM fit.}^2}(m_h,m_t,\alpha,\alpha_s)\left[1 + \frac{s_W^2}{c_W^2-s_W^2}\Delta
       r_{\text{SUSY}}^{(n)} \right]},\,
    \label{eq:MWfit}
\end{equation}
where $M_W^{\text{SM fit.}}$ is a numerical fit that incorporates the SM higher-order
corrections as a function of the SM input parameters. It is important to stress
that the implicit dependence of $\Delta_{\text{SM}}^{\text{lit.}} r$ on the
value of $M_W$,
which is correctly taken into account in \NMSSMCALC via
\cref{eq:MWcalcit}, is lost when using fit formulas for $M_W$. In \NMSSMTools
the $M_W$ dependence is partially restored by determining $M_W^{\text{SM fit.}}$
from the fit formula given in \cite{Awramik:2003rn}, inverting
\cref{eq:MWcalcit} for $\Delta_{\text{SM}} r(M_W^{\text{SM fit.}})$ and adding the
$M_W$ dependence using a further fit formula,
\begin{subequations}
\begin{align}
    \Delta^{\text{\NMSSMTools}}r(M_W) =\,
         & \Delta_{\text{SUSY}}\, r(M_W) + \Delta_{\text{SM}}\, r(M_W^{\text{SM fit.}}) 
            + \sum_{n=1}^{3} a_n \left(M_W-M_W^{\text{SM fit.}}\right)^n\\
 \approx &\,   \Delta_{\text{SUSY}}\, r(M_W) + \Delta_{\text{SM}}^{\text{lit.}}r(M_W)\,.
\end{align}
\label{eq:MWnt}
\end{subequations}
We now discuss the different treatments of SUSY input parameters.
All (SM and BSM) quantities entering \cref{eq:MWfit} are defined in the
$\MSbar$/$\DRbar$ scheme. In \FlexibleSUSY, \cref{eq:MWfit} is evaluated with
all running parameters at $M_Z$ while in \SASP
it is evaluated using parameters defined at the SUSY input scale. Thus, \SASP is closer to the
approach of \NMSSMCALC and \NMSSMTools which compute $M_W$ using
the running SUSY input parameters that are
given at the SUSY input scale $\msusy^2=m_{\tilde{t}_{R}} m_{\tilde{Q}_3}$.
In order to account for this systematic difference in the
$M_W$ calculation, we modified the spectrum-generator generated by \FlexibleSUSY
to compute $M_W$ and $a_\mu$ at $\msusy$ rather than $M_Z$.
Likewise, $\SASP$ computes $a_\mu$ by default at $M_Z$ which we changed to
$\msusy$.
While \NMSSMCALC is in principle able to renormalize parts of the SUSY sector
OS, a consistent comparison among the different tools is easiest when
interpreting all SUSY parameters as $\DRbar$ parameters defined at $\msusy$. In particular, the
parameter $\text{Re} A_\lambda$ is used as input parameter in the $\DRbar$ scheme, in
contrast to the previous sections. Likewise, the renormalization of
the top/stop sector for the calculation of the Higgs boson masses is
performed in the $\DRbar$ scheme. The $M_W$ prediction in \NMSSMCALC, however, is still
carried out in the OS scheme for the SM-sector described in \cref{sec:WMass}.
\NMSSMTools interprets $\tan\beta$ per default to be defined at $M_Z$ rather than
$\msusy$. Thus, we run the $\tan\beta(\msusy)$
used in the other codes to $M_Z$ (using two-loop RGEs generated with {\tt SARAH})
and use $\tan\beta(M_Z)$ in \NMSSMTools.
Furthermore, many of the involved codes also compute loop-corrected masses to
sfermions and electroweakinos which in turn are used in the $M_W$ prediction. We
also disabled such calculations in all programs as far as possible.
Another important ingredient for the $W$ mass prediction in supersymmetric
theories is the Higgs mass prediction.
For a detailed discussion of the different ingredients to the NMSSM Higgs mass
prediction as well as comparisons between the various spectrum generators we refer to \cite{Staub:2015aea,Slavich:2020zjv}.
In the context of the $W$-mass prediction, a common approach is to use the
loop-corrected Higgs mass, which
is around \unit[125]{GeV} in phenomenologically viable scenarios,
in the SM-part of the $M_W$ calculation. This ensures that, for a parameter point with
$m_h^{\text{(loop)}}\approx\unit[125]{GeV}$, the NMSSM in the decoupling limit predicts
the same numerical value for $M_W$ as the SM.
This is the approach implemented
in \FlexibleSUSY, \NMSSMCALC and \SASP. \NMSSMTools uses a fixed value of
$m_h=\unit[125.2]{GeV}$ in the SM fit formula. Furthermore, the SM-like Higgs
boson is not necessarily the lightest scalar in the spectrum since the
singlet-like states can in principle be lighter. For this reason, \NMSSMCALC
automatically determines the SM-like Higgs boson (based on the structure of the
mixing matrix) which is to be used in the SM part of the calculation. In case of
\FlexibleSUSY, this information can be given via the {\tt SLHA} input file by
the user.
To our best knowledge, \SASP does not have a mechanism to determine the
SM-like Higgs boson in the $M_W$ calculation but always assumes it to be the
lightest scalar state. Since we perform the comparison between the different
programs using a parameter point which has a
singlet state lighter than \unit[125]{GeV}, we modified the {\tt SARAH}
generated {\tt SPheno} code such that is also takes the index of the SM-like
Higgs boson as additional input in the {\tt SLHA} input file.

By construction the different methodologies to determine $M_W$ described by
\cref{eq:MWcalcit}, \cref{eq:MWfit} and \cref{eq:MWnt}, yield the precise SM result
in the limit of heavy superpartners (see the discussion in \cite{Athron:2022isz} for more details on the correct
decoupling behaviour).
We explicitly verified that all codes still feature a
proper decoupling behaviour after we applied the changes discussed above.

\begin{figure}[tb]
    \centering
        \includegraphics[width=0.99\textwidth]{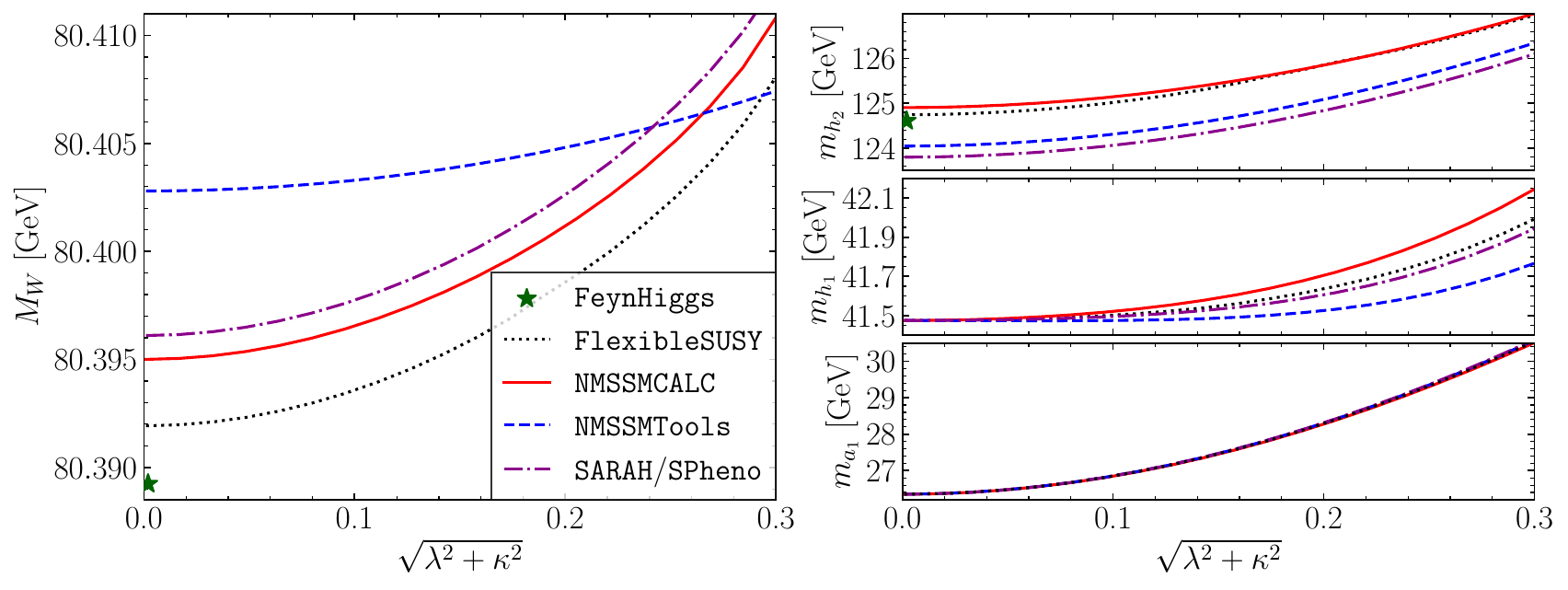}
        \caption{Comparison between \NMSSMCALC (red solid), \NMSSMTools (blue
            dashed), \SASP (magenta dash-dotted) and \FlexibleSUSY (black dotted)
            for the parameter point {\tt BP3} as a function of
            $\sqrt{\lambda^2+\kappa^2}$. Left: $W$-mass prediction. Right (up to
            down): prediction of the SM-like Higgs boson mass $m_{h_2}$, the
            lightest CP-even Higgs mass $M_{h_1}$ and lightest CP-odd mass
            $m_{a_1}$, respectively.
        } 
    \label{fig:toolcomparisonBP3}
\end{figure}

In order to compare the $m_h$, $M_W$ and $a_\mu$ prediction
numerically between the four codes \FlexibleSUSY~{\tt 2.7.1}, \NMSSMCALC~{\tt
5.2.0}, \NMSSMTools~{\tt 6.0.0} and {\tt SARAH 4.15.1}, we chose
two parameter benchmark points. The first point, {\tt P1}, was already
discussed in \cref{eq:paramP1} and features rather light electroweakinos.
The second, {\tt BP3}, was introduced in
\cite{Domingo:2022pde}\footnote{Note that {\tt BP3} has no preferred
features compared to the other parameter points ({\tt BP1}, {\tt BP2}, {\tt BP4}) in
\cite{Domingo:2022pde} but was chosen by matter of taste.}.
This parameter point is characterized by large one-loop corrections to
the $W$ boson mass due to very light sleptons with
masses of $\order{\unit[100]{GeV}}$.
Another interesting feature of this point is that the singlet-like CP-even and -odd Higgs bosons
have masses lighter than \unit[50]{GeV}. Therefore, both chosen
parameter points give the opportunity to compare the four considered $M_W$ codes under rather
extreme conditions.
For convenience, we reprint the input parameters for {\tt BP3} from
\cite{Domingo:2022pde} using the notation introduced in
\cref{sec:tree-levelspectrum},
\beq
&\text{\tt BP3}:\quad&  
m_{\tilde{t}_R}=2144\,\gev \,,\;
 m_{\tilde{Q}_3}=1112\,\gev\,,\; m_{\tilde{b}_R}=1539\,\gev\,, \; \non\\ 
 && 
m_{\tilde{L}_{1,2}}= 131.9\,\gev\,,\; m_{\tilde{e}_R,\tilde{\mu}_R}=103.6\,\gev \,,\; 
 \non\\
&& 
m_{\tilde{L}_{3}}= 205.2\,\gev\,,\; m_{\tilde{\tau}_R}=238.6\,\gev \,,\; |A_{u,c,t}| = 3971.2\,\gev\, ,\;
 \non\\ 
&&  |A_{d,s,b}|=1210.3\,\gev\,,\;
|A_{e,\mu}| =3643\,\gev\,,\; |A_{\tau}| =2052.4\,\gev\,,\; \label{eq:paramBP3} \\ \non
&& |M_1| = 178.3\,\gev,\; |M_2|= 128.6\,\gev\,,\; |M_3|=1757.6\,\gev \,,\\ \non
&& \lambda = 0.1229\,\gev,\; \kappa= 0.0128\,\gev\,,\; \tan\beta=8.7199\,\gev \,,\\ \non
&& \mu_{\text{eff}} = 212\,\gev,\; \text{Re}A_\kappa= -10.48.\,\gev\,,\;  \text{Re}A_\lambda= 2245\,\gev\,, \\ \non
&&  \varphi_{A_{e,\mu,\tau}}=0\, ,\; \varphi_{A_{u,c,t}}=\pi\,,\; 
\varphi_{A_{d,s,b}}=\varphi_{M_1}=\varphi_{M_2}=\varphi_{M_3}=0
 \;.
\eeq
In \cref{tab:comparison} we compare the prediction for $m_h$, $M_W$ and
$a_\mu$ obtained with all four codes for {\tt P1} and {\tt BP3}.
Additionally, we list the uncertainty estimates for $M_W$ and $a_\mu$ that are
returned by the programs \NMSSMTools and \FlexibleSUSY. The uncertainty for
$M_W$ in \NMSSMTools consists of a parametric uncertainty which is dominated by
the top quark mass (\cf \cref{sec:MWuncert}), the uncertainty due to the use of
the fit formula in \cref{eq:MWnt} and a SUSY uncertainty of about \unit[5]{MeV}. The uncertainty for
$m_h$ due to missing higher-orders can be estimated to be at least about $\gtrsim\unit[1]{GeV}$
\cite{Dao:2019qaz,Slavich:2020zjv,Dao:2021khm}.
Since the parameter point {\tt P1} is now interpreted with stop masses and
stop-Higgs trilinear couplings defined in the $\DRbar$ scheme, rather than OS,
the obtained Higgs boson mass for {\tt P1} is no longer at \unit[125]{GeV} but
around $m_h\approx\unit[119]{GeV}$.
Despite the fact that there are
many differences in the treatment of the parameters
between the four programs, the obtained results for the loop-corrected 
SM-like Higgs boson mass, for the anomalous magnetic
moment $a_{\mu}$ and for $M_W$ are overall in good agreement.
In particular the $M_W$ prediction is in agreement between all four codes even
if we only impose an uncertainty of $\unit[4-7]{MeV}$ which is about the size of
the $M_W$ uncertainty of the SM prediction, \cf \cref{sec:MWuncert}. The result for $a_\mu$
obtained with \NMSSMCALC for {\tt P1} is far outside the claimed uncertainty of \NMSSMTools and
\FlexibleSUSY, which were obtained by changing the scale at which $a_\mu$ is
calculated between $\msusy/2$ and $2\msusy$. This is due to the
scale-choice used in the calculation of $a_\mu$ which is fixed
in \NMSSMCALC to $\mu^2=m_{\tilde{L}_2}m_{\tilde{R}_2}$ while in the other
codes it is chosen dynamically to be the smallest mass of all sleptons and
electroweakinos. For {\tt P1}, the lightest SUSY particle is an electroweakino 
while for {\tt BP3} it is a slepton. Thus, we have better agreement for {\tt
BP3} than for {\tt P1}. We explicitly verified that \NMSSMCALC predicts a
similar number for {\tt P1} of $a_\mu\approx 3\times 10^{-10}$ when using the
lightest neutralino mass as the renormalisation scale.
\begin{table}[b]
\begin{center}
\begin{tabular}{|c||c||c|c|c|c|}
\hline
              &                        & \FlexibleSUSY  & \NMSSMCALC & \NMSSMTools     & \SASP     \\ \hline \hline
    {\tt P1}  & $m_{h}$ [GeV]          & 119.77         & 119.19     & 118.61          & 118.95    \\ \hline
              & $M_W$  [MeV]           & 80366.3        & 80365.7    & 80370.8$\pm23$  & 80366.2   \\ \hline
              & $a_\mu \times 10^{9}$  & 0.29$\pm0.01$  & 0.256      & 0.329$\pm0.03$  & 0.33      \\ \hline  \hline
    {\tt BP3} & $m_{h}$ [GeV]          & 125.60         & 125.63     & 124.63          & 123.97    \\ \hline
              & $M_W$ [MeV]            & 80396.9        & 80400.0    & 80404.2$\pm22$  & 80401.3   \\ \hline 
              & $a_\mu \times 10^{9}$  & 2.98$\pm0.45$  & 2.89       & 3.19$\pm 0.34$  & 3.70      \\ \hline 
\end{tabular}
\caption{Comparison of the prediction for the SM-like Higgs boson mass, the $W$
boson mass and the muon anomalous magnetic moment using \FlexibleSUSY,
\NMSSMCALC, \NMSSMTools, and \SASP. The benchmark point {\tt P1}, which was obtained from a scan in \NMSSMCALC using the OS scheme, is now interpreted in the $\DRbar$ scheme (which is why the Higgs mass is no longer at \unit[125]{GeV}).} 
\label{tab:comparison} 
\end{center}
\end{table}   
\begin{figure}[tb]
    \centering
        \includegraphics[width=0.99\textwidth]{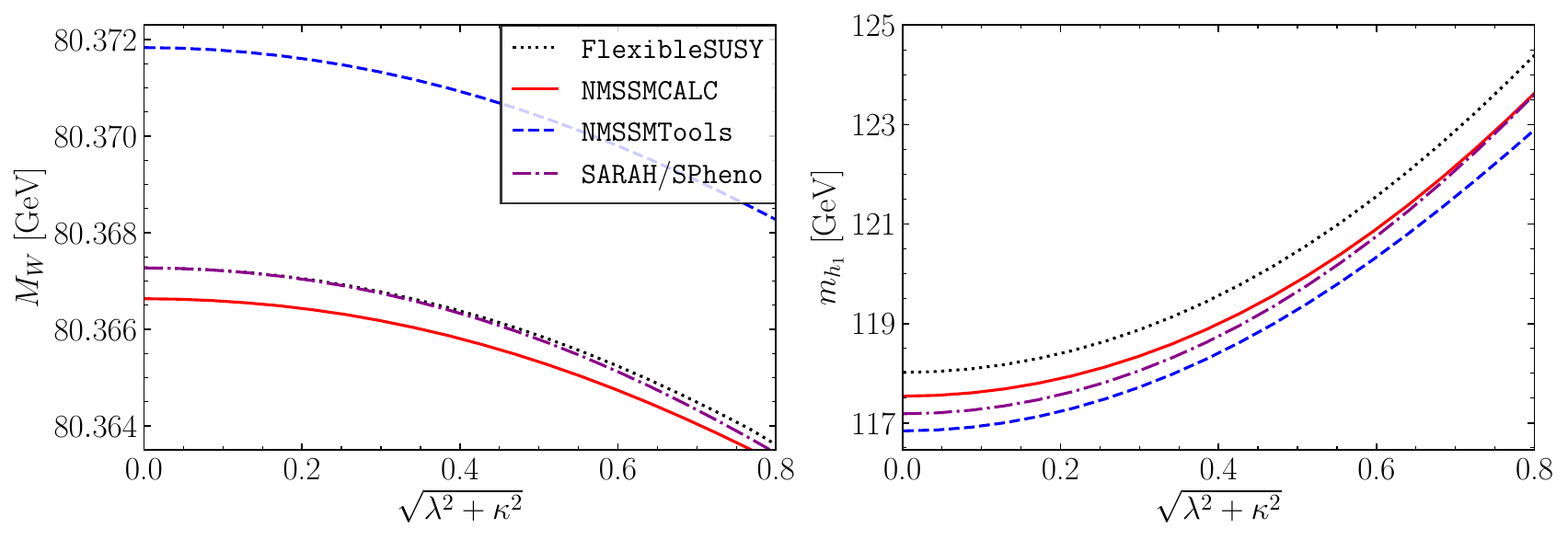}
        \caption{Comparison between \NMSSMCALC (red solid), \NMSSMTools (blue
            dashed), \SASP (magenta dash-dotted) and \FlexibleSUSY (black dotted)
            for the parameter point {\tt P1} in the $\DRbar$ scheme as a function of
            $\sqrt{\lambda^2+\kappa^2}$.
             Left: $W$-mass prediction. Right: prediction of the SM-like Higgs
             boson mass $m_{h_1}$.
        } 
    \label{fig:toolcomparisonBP1}
\end{figure}
In the following we compare the dependence of the $M_W$ prediction onto the superpotential
parameters $\lambda$ and $\kappa$ between the different codes. In
\cref{fig:toolcomparisonBP1} (left) we plot the $M_W$ prediction
obtained with the four codes. In the right plot we show the obtained value for
$m_h$ to validate if similar Higgs mass values are used in the respective SM
higher-order results.
We observe that $M_W$ obtained with $\FlexibleSUSY$ and $\SASP$
agree almost perfectly and only start to slightly deviate for very large values
of $\slk\gtrsim 0.6$. The level of agreement in $M_W$, however, always need to
be seen in the light of the $m_h$ prediction. In case of \FlexibleSUSY and \SASP
the Higgs mass predictions differ by about \unit[0.8-1]{MeV}
which means that their SM-prediction actually differs by \unit[0.2-0.5]{MeV}
such that the perfect agreement in \cref{fig:toolcomparisonBP1} (left) seems
accidental. Furthermore, they never differ from the \NMSSMCALC prediction for $M_W$
by more than \unit[0.63]{MeV}. \NMSSMTools, however, seems to always predict a
$W$ mass which is about \unit[2]{MeV} larger even though its Higgs mass
prediction is also relatively close to the other codes.
We perform a similar analysis for the parameter point {\tt BP3} in
\cref{fig:toolcomparisonBP3}. For this point we plot, in addition to the SM-like
Higgs boson mass $m_{h_2}$, the mass of the
lightest CP-even and -odd Higgs boson, $m_{h_1}$ and $m_{a_1}$, respectively.
For all three shown scalar masses we find good agreement and a similar
behaviour in terms of $\slk$ which is rather flat in the shown range of
$\slk\leq 0.3$. For larger values of $\slk$ the singlet-like CP-even state
becomes tachyonic at tree-level.
The $M_W$ prediction of \FlexibleSUSY, \NMSSMCALC and \SASP shows almost exactly the 
same behaviour when varying $\slk$. The \NMSSMCALC $M_W$ prediction
differs with the one of \FlexibleSUSY (\SASP) by at most \unit[1.7]{MeV}
(\unit[3.1]{MeV}) which is smaller than the SM-uncertainty.
The prediction obtained with \NMSSMTools, however, seems to behave much flatter
for large values of $\slk$. This is likely because \NMSSMTools seems to use the
loop-corrected scalar masses in all parts of the $M_W$ calculation. Furthermore,
we find that the $M_W$ predicted by \NMSSMTools agrees much better with the other
codes in \cref{fig:toolcomparisonBP1,fig:toolcomparisonBP3}, if we remove the
$M_W$-restoring fit function in \cref{eq:MWnt} from its prediction.
Thus, we suspect that the fit coefficients in \cref{eq:MWnt} in \NMSSMTools are
outdated. Finally, in the MSSM limit $\slk\to 0$, we are also able to also
compare with the code {\tt FeynHiggs 2.19.0} 
which calculates $m_h$ and $M_W$ in the MSSM rather than the NMSSM. For a
detailed description of the $M_W$ calculation in {\tt FeynHiggs} we refer to
\cite{Heinemeyer:2013dia}.
In \cref{fig:toolcomparisonBP3} we
find that {\tt FeynHiggs} yields the smallest $M_W$ prediction which is,
however, still in good agreement with the other codes given the SM uncertainty
alone. In particular the difference to the \NMSSMCALC prediction is about
\unit[5.7]{MeV}.

\subsection{CP-Violating Effects in the $M_W$ Prediction}
\begin{figure}[t]
    \centering
        \includegraphics[width=0.9\textwidth]{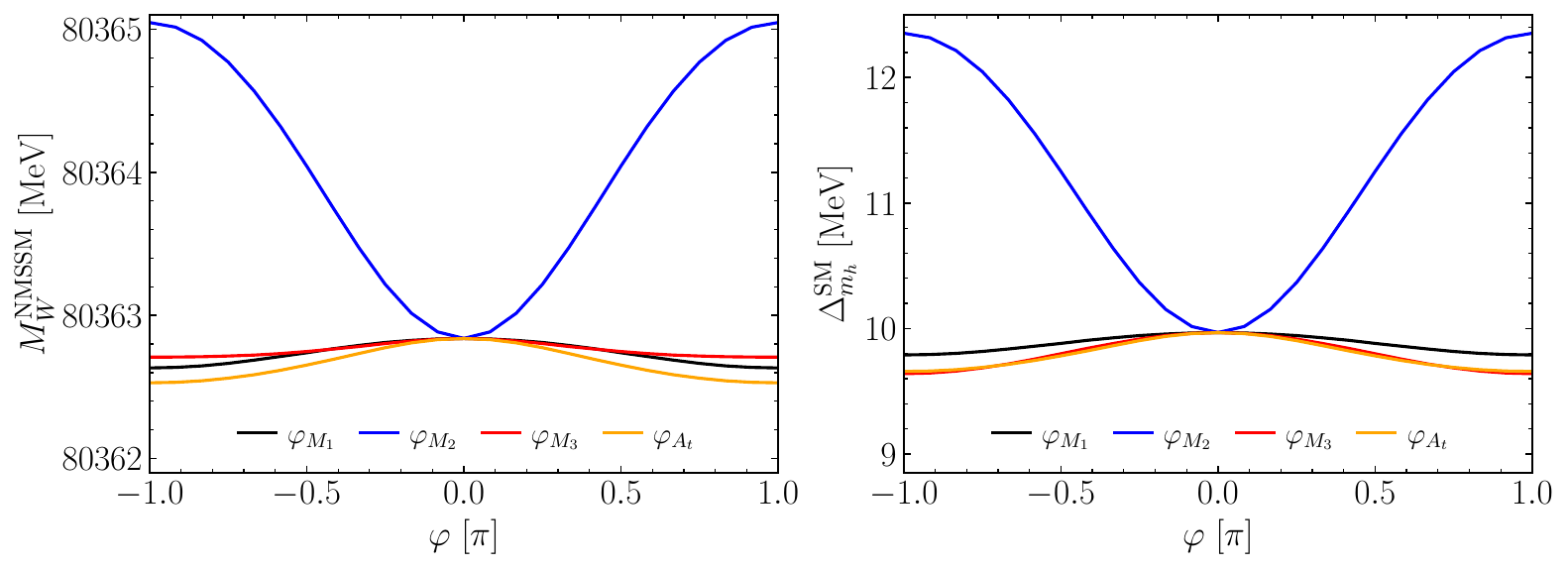}
    \caption{Left panel: $M_W$ at two-loop ${\cal
        O}(\alpha_{\text{new}}^2)$ as a
      function of the complex phases $\varphi_{M_1}$ (black),
      $\varphi_{M_2}$ (blue),  
    $\varphi_{M_3}$ (red) and $\varphi_{A_t}$  (orange) for the parameter point {\tt P1}.   
     Right panel: same as the left panel but showing the difference between the
 NMSSM and SM $M_W$ prediction that has been obtained using \NMSSMCALC and the
value of the SM-like Higgs mass prediction for the particular values of
$\varphi_i$, \cf \cref{eq:deltas}.
}
    \label{fig:P1-phase}
\end{figure}

In this section we study the effect of CP-violating beyond-the-SM phases on the $M_W$
calculation. We consider the parameter point {\tt P1}, which was initially defined in the CP-conserving NMSSM, and
study its dependence on the phases of $A_t,\, M_1,\, M_2,$ and
$M_3$. Note, that this investigation
is for illustrative purpose and we hence do not check the validity of
the phases w.r.t.~the EDM constraints.
Figure~\ref{fig:P1-phase} (left) shows the resulting prediction
for $M_W$ at two-loop ${\cal O}(\alpha_{\text{new}}^2)$ if the phases are varied
individually for the parameter point {\tt P1}. In the right plot we show the
difference of the $M_W$ prediction to the SM prediction $\deltasm$ defined in
\cref{eq:deltas}.
We find that the phase of $\varphi_{M_2}$ has the largest impact on $M_W$ for
this parameter point of about \unit[2]{MeV} while the overall SUSY corrections
$\deltasm$ are at most \unit[12]{MeV} which is due to the very light
electroweakino masses. The stop quarks and gluinos have already been decoupled
from the $M_W$ prediction as they are heavier than \unit[1]{TeV}. Thus, also the
phases $\varphi_{A_t}$ and $\varphi_{M_3}$ only impact $M_W$ at the sub-MeV
level, which is in agreement with the findings in \cite{Heinemeyer:2006px} for
the MSSM.
We furthermore observe that the phase dependence is dominating in the one-loop
corrections while the two-loop corrections only lead to a change in the phase
dependence at the sub-MeV level. In conclusion, the overall phase dependence is
smaller than the size of the total shift of the SUSY corrections to $M_W$.
\section{Conclusions}
\label{sec:conclusions}
In this paper, a consistent inclusion of the two-loop $\orderqcd$ and
$\ordernew$ corrections to the $\rho$ parameter and its application in the 
calculation of the $W$ boson mass has been presented in the context of the complex NMSSM.
Both corrections have been computed by our group in the previous computation of the loop-corrected Higgs boson masses
at the corresponding orders. These two calculations use the gaugeless limit and
the zero external momentum approximation. The renormalization features a mixed
OS/$\DRbar$ scheme and conveniently allows to switch between OS and $\DRbar$
conditions used in the top/stop sector and for the charged Higgs boson mass.
A~scheme change in the top/stop sector is used to estimate the uncertainty in
the $\rho$ parameter prediction due to missing higher-orders.
We showed that the $\orderqcd$ and $\ordernew$
corrections to the $\rho$ parameter are significant and can help to reduce
the theory uncertainty. After subtracting the SM corrections we add them back in
the evaluation of the $W$ mass including all known higher-order SM corrections in the OS scheme.
We show that the effects arising from $\orderqcd$ and
$\ordernew$ for $M_W$ are of the order of a few MeV which is smaller than the
parametric uncertainty of the top mass and is of similar size as the missing higher-order SM corrections.
We have implemented all corrections in the new version of the Fortran code {\tt NMSSMCALC} which takes
into account the most up-to-date higher-order
corrections for the CP-violating NMSSM in the computation of the Higgs boson masses, Higgs boson decay widths and branching ratios, the muon anomalous magnetic
moment $a_\mu$, electric dipole moments, and the $W$ boson mass.
Finally, we have performed a detailed comparison of the $W$ boson mass, Higgs
boson mass, and muon anomalous magnetic moment prediction between \NMSSMCALC and
the public spectrum generators \FlexibleSUSY, \SASP and \NMSSMTools. We found
good agreement between all calculations once different treatments of the
most-important input parameters and renormalization scales are taken into account.

\section*{Acknowledgements}
The authors thank G. Weiglein for various discussions about the SM higher-order
corrections.
The research of MM was supported by the Deutsche
Forschungsgemeinschaft (DFG, German Research Foundation) under grant
396021762 - TRR 257. T.N.D is funded by the Vietnam National Foundation for Science
and Technology Development (NAFOSTED) under grant number 103.01-2020.17.
M.G. acknowledges support by the Deutsche Forschungsgemeinschaft (DFG, German
Research Foundation) under Germany's Excellence Strategy --- EXC 2121 ``Quantum
Universe'' --- 390833306. This work has been partially funded by the Deutsche
Forschungsgemeinschaft (DFG, German Research Foundation) --- 491245950.

\bibliographystyle{JHEP}
\bibliography{paper}

\end{document}